\documentclass[useAMS,usenatbib,referee]{mnras}
\usepackage[dvipdfmx]{graphicx}
\usepackage{pdfpages}
\usepackage{amsmath}  % Advanced maths commands
\usepackage{amssymb}  % Extra maths symbols
\usepackage{multirow}
\newcommand{\etal }{{et al.} }

\newcommand{\msun}{{\rm M}_\odot}

\newcommand{\cc}{{\rm cm^{-3}}}
\newcommand{\dg}{^\circ}
\newcommand{\kms}{{\rm km\,s^{-1}}}

\newcommand{\vect}[1]{\mbox{\boldmath$#1$}}

\title
[High-velocity Jets and Low-velocity Flows]%[**/45 characters]
{Can High-velocity Protostellar Jets Help to Drive Low-velocity Outflow?}

\author[Machida]{Masahiro N. Machida$^{1}$
\\
$^{1}$
Department of Earth and Planetary Sciences, Faculty of Sciences, Kyushu University, Fukuoka, Fukuoka 819-0395, Japan\\
}
\date{Accepted XXX. Received YYY; in original form ZZZ}
\pubyear{2020}

\begin{document}
\label{firstpage}
\pagerange{\pageref{firstpage}--\pageref{lastpage}}
\maketitle

\begin{abstract}%[***/200 words for Letters]
Using three-dimensional magnetohydrodynamics simulations, the driving of protostellar jets is investigated in different star-forming cores with the parameters of magnetic field strength and  mass accretion rate. 
Powerful high-velocity jets appear in strongly magnetized clouds when the mass accretion rate onto the protostellar system is lower than $\dot{M} \lesssim 10^{-3}\,\msun$\,yr$^{-1}$. 
On the other hand,  even at this mass accretion rate range,  no jets appear for magnetic fields of prestellar  clouds as weak as $\mu_0 \gtrsim 5$--$10$, where $\mu_0$ is the mass-to-flux ratio normalized by the critical value $(2\pi G^{1/2})^{-1}$. 
For  $\dot{M}\gtrsim 10^{-3}\,\msun$\,yr$^{-1}$, although jets usually appear just after protostar formation independent of the magnetic field strength, they soon weaken and finally disappear. 
Thus, they cannot help drive the low-velocity outflow when there is no low-velocity flow just before protostar formation.
As a result,  no significant mass ejection occurs during the early mass accretion phase  either when the prestellar cloud is weaky magnetized or when the mass accretion rate is very high.
Thus, protostars formed in such environments would trace different evolutionary paths from the normal star formation process.    
	\end{abstract}
%%%%%%%%%%%%%%%%%%%%%%%%%%%%%%%%%%%%%%%%
\section{Introduction}
\label{sec:intro}
Protostellar outflows (low-velocity wide-angle outflows and high-velocity collimated jets)  are ubiquitously observed in various star-forming regions \citep{wu04}.  
They are considered to universally appear in the star formation process.
Outflows are powered by accreting matter, in which the gravitational energy of the accreting matter is converted into the outflow kinetic energy through the magnetic field  (or via magnetic energy; \citealt{uchida85,blandford82,tomisaka02}). 
Past studies have shown  that protostellar outflow is necessary for removing the angular momentum of star-forming clouds \citep{tomisaka00,joos12,masson16,hirano20}. 
In addition, the star formation efficiency should be related to the mass ejection rate of  the outflow, as a large fraction of the cloud mass is ejected by the protostellar outflow \citep{matzner00,arce07,machida13}. 
Thus, the protostellar outflow plays significant roles  in the star formation process \citep{tanaka17,tanaka18}. 
Nevertheless, some objects (or protostars) have been observed to show no sign of outflow \citep[e.g.][]{tokuda18,aso19}.
%% namely protostars embedded in a dense infalling envelope and thus in the main accretion phase. 

In previous studies \citep[][hereafter Papers I, II and III]{matsushita17,matsushita18,machida20}, we investigated the driving condition of low-velocity outflows in various star-forming cloud cores.  
After performing long-term simulations, we concluded that low-velocity outflow fails to appear or is delayed  when the magnetic field of the star-forming core is weak and/or when the mass accretion rate onto the central region is high (see also \citealt{wurster16} and \citealt{lewis17}). 
We called  these cases  ``failed or delayed outflows'' and showed that they correspond to the objects that were observed to show no protostellar outflow \citep{aso19}. 
For the failed and delayed outflow cases, a high ram pressure of the infalling envelope suppresses the outflow driving (see, Papers I and III). 
In addition, for the delayed outflow case, the outflow begins to evolve only after a large part of the infalling envelope dissipates, because the ram pressure of the envelope weakens as the density of the infalling envelope becomes low (Paper III).
However, there has been criticism about the numerical settings regarding the failed and delayed outflow cases.
In our previous studies (Papers I--III), we did not resolve the protostar itself. 
Instead of resolving the protostar, we used sink cells which cover both the inner disk region and protostar. 
Although the sink cell method can accelerate the time evolution of star-forming clouds in the simulation,  the region around the protostar cannot be resolved \citep{machida10}. 
Thus, we could not simulate a high-velocity outflow component (i.e. high-velocity jet) driven near the protostar. 
Note that high-velocity components can be seen in the high-spatial resolution simulations of low-mass star formation \citep[][]{tomisaka02,banerjee06,machida06,tomida13,bate14,machida14,lewis17,vaytet18,machida19}.
Note also that few studies focused on the outflow in the  high-mass star formation simulations with sink cells, in which the flows driven near the protostar were not spatially resolved \citep{seifried11, seifried12, kolligan18}.

In both observational and theoretical works, it is well known that two types of flow appear in the star formation process \citep[e.g.][]{arce07,machida08}. 
A high-velocity collimated flow (so-called high-velocity jets) appears near the protostar, while a low-velocity wide-angle flow (so-called low-velocity outflow) appears in the region far from the protostar or in the outer edge of the disk \citep[see review by][]{inutsuka12}.  
Note that no flow appears in the intermediate disk region because the magnetic field is not well coupled with neutrals \citep{machida07,xu21}. 
In our previous studies (Papers I--III), we did not resolve the high-velocity jets.
Thus, if the protostar and the disk inner region  are spatially resolved, high-velocity jets possibly appear and may help to drive a wide-angle flow (or low-velocity outflow). 
In other words, even if the low-velocity outflow does not appear in a simulation using sink cells, it may appear with help from high-velocity jets in high-resolution simulations without sink cells.  
In addition, there is a long-standing debate about driving wide-angle low-velocity outflows. 
There are two conflicting models or scenarios for driving low-velocity outflow, the entrainment and direct-driven scenarios. 
The low-velocity outflow is entrained by the high-velocity jets in the former model \citep{arce07}, while the low-velocity outflow is  driven directly from the outer disk region in the latter \citep{matsushita19}. 
Although both recent observations and numerical simulations support the direct driven scenario \citep{hennebelle08,machida14,bjerkeli16,hirota17,alves17,tabone18,devalon20,lee21,marchand20,wurster21}, entrainment may not be completely rejected. 
It may be  possible for the high-velocity jets to entrain a part of the infalling gas and produce  the low-velocity component.

In Paper III using simulations with sink cells, I showed that  the low-velocity outflow fails to appear or appears in a late stage in weakly magnetized clouds when the mass accretion rate onto the central region is high.
In this study, I investigate whether the high-velocity flow can appear and contribute to driving the low-velocity outflow with almost the  same settings  as in Paper III but without using sink cells, in which the protostar is spatially resolved with a spatial resolution of $7.9\times10^{-3}$\,au.
%%Therefore, this study is complementary to our previous works of Papers I-- III.

The structure of this paper is as follows.
The numerical settings and methods are described in \S2 and the calculation results are presented in \S3.
I discuss the effect of the high-velocity jets on the low-velocity outflow in \S4. 
A summary is presented in \S5.

%%%%%%%%%%%%%
%%% Table1%%%
%%%%%%%%%%%%%
\renewcommand{\arraystretch}{1.2}
\begin{table}
\begin{center}
\begin{tabular}{c||cccccccccc} \hline
Model & $f$ &  $\mu_0$ & $B_0$ [G] & $\Omega_0$ [s$^{-1}$] & $\alpha_0$ & $\beta_0$ & $\gamma_0$  & $M_{\rm cl}$ \, [$\msun$]  &  Jet & Outflow\\
\hline
AM3   & 1.4 & 3  & $1.0\times10^{-5}$ &  $3.2\times10^{-14}$   & 0.5 & 0.02 & 0.083 & 11 &  S & S\\
AM5   & 1.4 & 5  & $6.2\times10^{-6}$ & $3.2\times10^{-14}$  & 0.5 &  0.02& 0.030 & 11 & N & D\\
AM10 & 1.4 & 10  & $3.1\times10^{-6}$ & $3.2\times10^{-14}$  & 0.5 & 0.02 & $7.5\times10^{-3}$ & 11 & N & D\\
AM20 & 1.4 & 20  & $1.6\times10^{-6}$ & $3.2\times10^{-14}$  & 0.5 &  0.02& $1.9\times10^{-3}$ & 11 & N & D\\
\hline
BM3   & 3.4 & 3  & $2.5\times10^{-5}$ & $5.0\times10^{-14}$  & 0.2 & 0.02& 0.083 & 28  & S & S \\
BM5   & 3.4 & 5  & $1.5\times10^{-5}$ & $5.0\times10^{-14}$  & 0.2 & 0.02& 0.030 & 28  & N & D \\
BM10 & 3.4 & 10  & $7.5\times10^{-6}$ & $5.0\times10^{-14}$  & 0.2 & 0.02& $7.5\times10^{-3}$ & 28 & N & F \\
BM20 & 3.4 & 20  & $3.7\times10^{-6}$ & $5.0\times10^{-14}$  & 0.2 & 0.02& $1.9\times10^{-3}$ & 28 & N & F\\
\hline
CM3   & 8.4 & 3  & $6.2\times10^{-5}$ & $7.9\times10^{-14}$  & 0.08 &  0.02& 0.083  & 68 & S  & S\\
CM5   & 8.4 & 5  & $3.7\times10^{-5}$ & $7.9\times10^{-14}$  & 0.08 &  0.02& 0.030  & 68 & W & F\\
CM10 & 8.4 & 10  & $1.9\times10^{-5}$ & $7.9\times10^{-14}$  & 0.08 &  0.02& $7.5\times10^{-3}$  & 68 & W & F \\
CM20 & 8.4 & 20  & $6.2\times10^{-6}$ & $7.9\times10^{-14}$  & 0.08 &  0.02& $1.9\times10^{-3}$  & 68 & W & F \\
\hline
DM3 & 16.8 & 3 & $1.2\times10^{-4}$ & $1.1\times10^{-13}$ & 0.04 &  0.02& 0.083  & 132 & S & S\\
DM5 & 16.8 & 5 & $7.4\times10^{-5}$ & $1.1\times10^{-13}$ & 0.04 &  0.02& 0.030  & 132 & W & F \\
DM10 & 16.8 & 10 & $3.7\times10^{-5}$ & $1.1\times10^{-13}$ & 0.04 & 0.02& $7.5\times10^{-3}$  & 132 & W & F\\
DM20 & 16.8 & 20  & $1.8\times10^{-5}$ & $1.1\times10^{-13}$  & 0.04 &  0.02& $1.9\times10^{-3}$  & 132 & W & F \\
\hline
EM3 & 33.6 & 3  & $2.4\times10^{-4}$ & $1.5\times10^{-13}$  & 0.02 &  0.02& 0.030 & 272  & N & S\\
EM5 & 33.6 & 5 & $1.5\times10^{-4}$ & $1.5\times10^{-13}$  & 0.02 &  0.02& 0.083 & 272  & W & D\\
EM10 & 33.6 & 10  & $7.4\times10^{-5}$ & $1.5\times10^{-13}$  & 0.02 & 0.02 & $7.5 \times 10^{-3}$ & 272 & N & F \\
\hline
FM3  & 67.2 & 3  & $5.0\times10^{-4}$ & $2.2\times10^{-13}$  & 0.01 &  0.02& 0.083 & 545 & N & D\\
FM5  & 67.2 & 5  & $3.0\times10^{-4}$ & $2.2\times10^{-13}$  & 0.01 &  0.02& 0.030 & 545 & N & D\\
FM10 & 67.2 & 10  & $1.5\times10^{-5}$ & $2.2\times10^{-13}$  & 0.01 &  0.02& $7.5\times10^{-3}$  & 545 & W & D \\
\hline
\end{tabular}
\end{center}
\caption{
Model name, initial cloud parameters and calculation results.
Column 1 gives the model name. 
Columns 2 and 3 give the parameters $f$ and $\mu_0$. 
Columns 4 and 5 give the magnetic field strength $B_0$ and angular velocity $\Omega_0$ for the initial state. 
Columns 6--8 give the ratios of the thermal $\alpha_0$, rotational $\beta_0$ and magnetic $\gamma_0$ energies to the gravitational energy of the initial cloud. 
Column 9 gives the initial cloud mass.
Column 10 describes the calculation results of this study, in which `S', `W' and `N' means that a strong, weak and no jet appears,  respectively. 
Column 11 describes the calculation results of Paper III, in which `S', `D' and `F' mean that an outflow successfully appears, a delayed outflow appears and no (or failed) outflow appears, respectively. 
}
\label{table:1}
\end{table}

\section{Initial Conditions and Numerical Settings}
\label{sec:settings}
The initial conditions and numerical settings are almost the same as in Papers I--III.
Thus, I only briefly explain them in this section.
As the initial state, I adopt a spherical cloud core with a Bonnor--Ebert  (B.E.) density profile with a central density of  $n_{\rm c,0}=10^5\, \cc$ and an isothermal temperature of $T_{\rm iso,0}=20$\,K.
The initial cloud has twice the critical B.E. radius ($R_{\rm cl}=4.1\times10^4$\,au).
The cloud density is enhanced by a factor of $f$ to promote gravitational collapse, in which $f$ is the density enhancement factor (Papers I and III) and is related to the cloud (thermal) stability $\alpha_0$ (ratio of thermal to gravitational energy).
Therefore, the initial cloud has a central density of $n_{\rm cl}= f \times 10^5\,\cc$ ($= f \times n_{\rm c,0}$). 
A uniform density $n_{\rm ISM}=1.25\times10^{-2}\, n_{\rm cl}$ is set outside the initial cloud ($r>R_{\rm cl}$).

A uniform magnetic field $B_0$ is imposed over the whole computational domain.
The magnetic field direction is set to be parallel to the $z$-axis.
A rigid rotation $\Omega_0$ is adopted within the cloud ($r<R_{\rm cl}$) and the rotation axis is set to be inclined by $\theta_0$ toward the $x$-axis from  the $z$-axis. 
Thus, the initial magnetic field is misaligned with the initial rotation axis.
On the other hand, an initial magnetic field parallel to the initial rotation axis was adopted in Papers I--III.
Recently, we have shown that the aligned cases are not very realistic \citep{hirano19,hirano20,machida20b}, because the disk normal, the propagation direction of the outflow  and jets and the angular momentum vector of  the disk and protostar are well  aligned, and the angular momentum in such systems is transported only in specific directions \citep[for details, see][]{ciardi10, joos12,hirano20}. 
Thus, the misaligned case should be  adopted as a general (or realistic) case. 
In this study, to limit the number of models, referring to the results in \citet{machida20b}, I fixed the angle $\theta_0$ to be $\theta_0=20\dg$, which can yield  strong outflow and jets in the low-mass star-formation process \citep{lewis15,machida20b}.

As the initial state, a total of 22 different prestellar clouds are prepared (Table~\ref{table:1}), in which the density enhancement factor $f$ (or $\alpha_0$) and the normalized mass-to-flux ratio $\mu_0$ (or magnetic field strength $B_0$) are parameters.
The parameter $\alpha_0$ (or $f$) controls the mass accretion rate onto the central region $\dot{M}$, in which the mass accretion rate is proportional to $\alpha_0^{-3/2}$ (for details, see Paper I).
I will detail the mass accretion rate for some of the models in \S\ref{sec:strong}. 
In this series of studies,  the mass accretion rate  parameter (or $\alpha_0$) determines whether star in formation is low- or high-mass.  
The normalized mass-to-flux ratio is defined as 
\begin{equation}
\mu_0 \equiv \left( \frac{M_{\rm cl}}{\Phi_{\rm cl}} \right) / \left( \frac{1}{2\pi G^{1/2}} \right), 
\end{equation}
where $\Phi_{\rm cl}$ is the magnetic flux of the initial cloud.
The initial magnetic field strength $B_0$ is adjusted so that the mass-to-flux ratio becomes $\mu_0=3$,  5, 10 and 20 in each initial cloud. 
The rotation rate $\Omega_0$ is determined to give $\beta_0=0.02$ in each model, where $\beta_0$ is the ratio of the rotational to gravitational energy.
The model name and cloud parameters are described  in Table~\ref{table:1}
\footnote{
I do not include models EM20 and FM20 to further limit the number of models. 
Since these models have a high mass accretion rate and high infall velocity, a very long calculation time is necessary to complete the calculation. 
}.
Except for the angle $\theta_0$ between the initial magnetic field direction and the rotation axis, the initial clouds are completely the same as in Paper III.

The numerical settings are almost the same as in Papers I--III. 
The basic equations (resistive MHD equations) are described in equations~(1)--(4) of \citet{machida19} and equation~(1) of \citet{machida14}.
The nested grid code is used to cover both prestellar core and protostar  \citep[for details of the code, see][]{machida04,machida05a,machida10,machida13}. 
At the beginning of the calculation, five levels, $l$, of nested grids are prepared. 
Each grid is composed of (i, j, k) = (64, 64, 64).
The initial prestellar cloud is set in the fifth level of the grid ($l=5$). 
The first level of the grid has a box size of $L(l=1)=6.6\times10^5$\,au and a cell width of $h(l=1)=1.0\times10^4$\,au.  
Both the box size and cell width halve with each increment of grid level $l$. 
A new finer grid is generated to satisfy the Truelove condition, in which the Jeans length is resolved with at least 16 cells \citep{truelove97}. 
The maximum grid level is set to $l_{\rm max}=21$, which has a box size of $L(l=l_{\rm max})= 0.51$\,au and a cell width of $h(l=l_{\rm max})=7.9\times10^{-3}$\,au. 

For the purpose of reference, I describe the difference in the numerical settings between this study and Papers I--III. 
%% is the treatment of the protostar. 
In Papers I--III, the central region is covered by the sink cells with a spatial resolution of $h= 0.62$\,au and a sink radius of $r_{\rm sink}=1$\,au, while in this study the protostar is resolved with $h=7.9\times10^{-3}$\,au without using sink cells. 
The detailed settings can be confirmed in \citet{machida14} and \citet{machida15}.
A small time step is required with a high spatial resolution  to ensure the Courant, Friedrichs and Lewy condition.
Thus, the simulation could be executed for about 10,000\,yr after protostar formation in Paper III, while only for $\sim500$\,yr after protostar formation in this study.

\section{Results}
\label{sec:results}
The evolution of outflows and jets for the models listed in Table~\ref{table:1} are presented  in this section. 
As described in \S\ref{sec:settings}, two parameters, the mass-to-flux ratio $\mu_0$ (corresponding to magnetic field strength) and  thermal stability $\alpha_0$ (corresponding to mass accretion rate), are used to characterize the models. 
%%The parameters  are almost the same as in Paper III, while some models in Paper III (AM2, BM2, CM2, DM2, EM2, FM2, EM20, FM20) are excluded from this study to limit the number of models. 
%%Note that the model names used are the same as those in Paper III when the parameters $\alpha_0$  and $\mu_0$ are the same. 
This section first focusses on  the evolution of jets for models with strong magnetic fields (AM3, BM3, CM3, DM3, EM3, FM3).
Then,  the jet driving for all the models in Table~\ref{table:1} is shown. 

In this paper, as shown in previous studies \citep{tomisaka02,banerjee06,machida14,machida19}, two components, the low-velocity outflow and high-velocity jet, appear from different disk regions. 
However, it is very difficult  to distinguish them. 
Thus, I simply call all the outflowing gas components including the low-velocity outflow and high-velocity jet the `outflow' in this section. 
In addition, I describe the high-velocity outflow component as `the high-velocity jet (or high-velocity component)', and the low-velocity outflow as `the low-velocity outflow (or low-velocity component)', though there is no clear velocity boundary between them.
It should be noted that, in Papers I--III, since the high-velocity component was not resolved and  does not appear, the low-velocity flow (i.e. all the outflowing gas) was referred to as  the outflow.

%%%%%%
% Fig. 1
%%%%%%
\begin{figure*}
\begin{center}
\includegraphics[width=0.8\columnwidth]{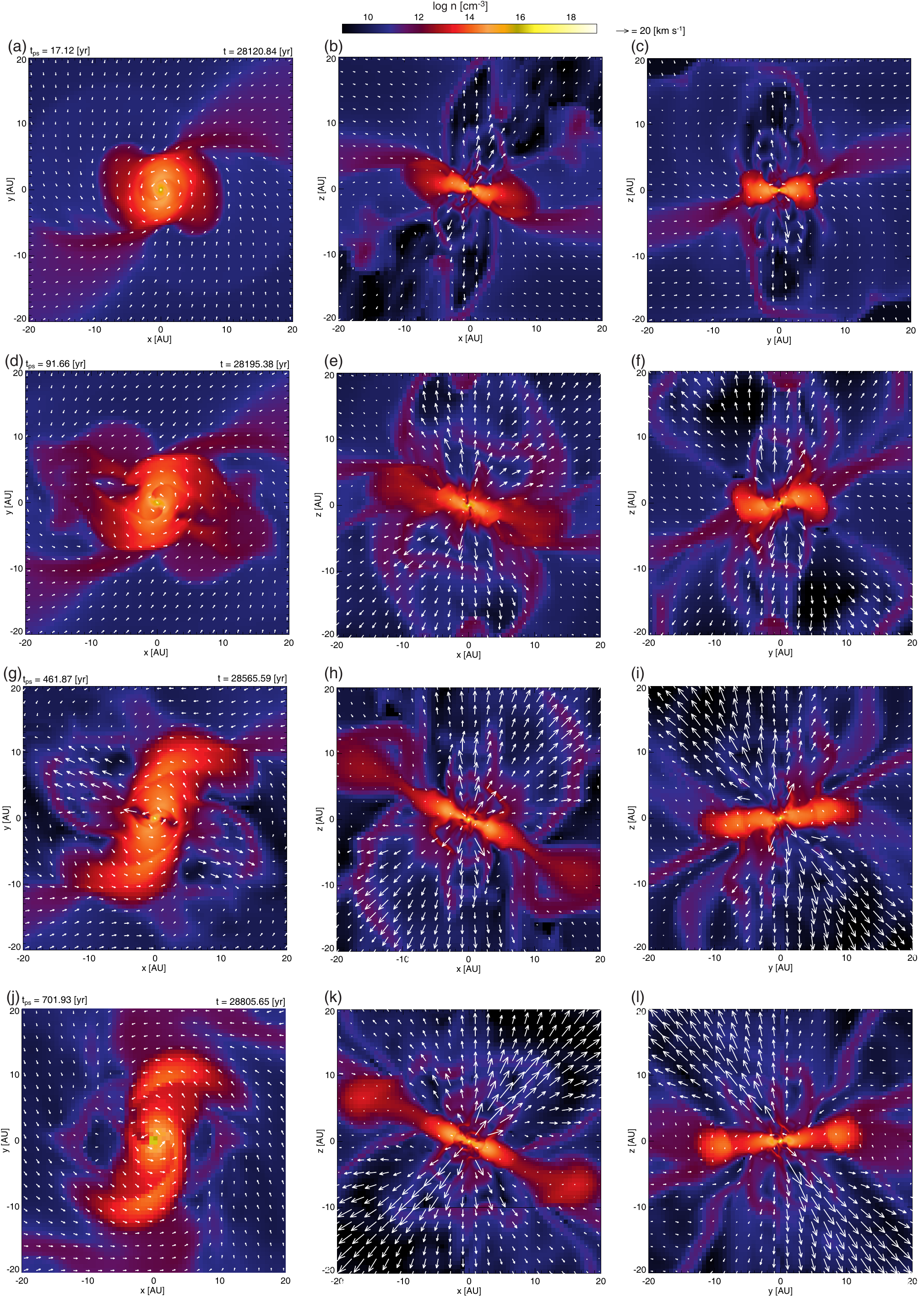}
\end{center}
\caption{
Density (color) and velocity (arrows) distributions at different epochs on the $z=0$ (left panels), $y=0$ (middle panels) and $x=0$ (right panels) planes for DM3 (see also DM3.avi). 
The elapsed time $t_{\rm ps}$ after protostar formation and that $t$ after the beginning of the cloud collapse are described in each left panel. 
}
\label{fig:1}
\end{figure*}

\subsection{Outflow Evolution for Strongly Magnetic Field Models} 
\label{sec:strong}
%%When the initial magnetic strength is as strong as $\mu_0=2$,  all the models (AM2, BM2, CM2, DM2, EM2, FM2) show a powerful outflow in Paper III. 
%%Meanwhile, among the models with $\mu_0=3$ (AM3, BM3, CM3, DM3, EM3, FM3), only FM3 with $\alpha_0=0.01$ shows no outflow just after protostar formation.
%%For FM3, the outflow begins to evolve after the infalling envelope dissipates (for details, see Paper III). 
%%Note that the mass accretion rate for FM3 is the highest among the models with $\mu_0=3$. 
%%The difference in magnetic field strength between FM2 ($B_0=7.4\times10^{-4}$\,G)  and FM3 ($B_0=5.0\times10^{-4}$\,G) is only a factor of 1.48. 
%%In Paper III, we concluded that a slight difference in magnetic field strength causes a significant difference in the star formation process ({\bf i.e. whether the outflow appears or not}). 
{\bf
%%Note that  the magnetic field strength required for the appearance of a clear outflow depends on the mass accretion rate. 
%%A clear outflow appears  with $\mu_0 < 5 $ when the mass accretion rate is lower than $\dot{M}\lesssim 7\times10^{-3}\,\msun$\,yr$^{-1}$, while it appears only with $\mu_0=2$ when $\dot{M}\gtrsim  7\times10^{-3}\, \msun$\,yr$^{-1}$ (for details, see Papers I and III).
}
%%As described above, since the high-velocity component (high-velocity jet) was ignored in Paper III, I investigate the effects on it here. 

Firstly, the cloud evolutions for DM3, EM3 and FM3 are shown. 
Figure~\ref{fig:1} shows the time sequence of the central region around the protostar for DM3 with the density and velocity distributions on the $z=0$ (left), $y=0$ (middle) and $x=0$ (right) planes plotted. 
DM3 has the parameters $\mu_0=3$ and $\alpha_0=0.04$.
We can confirm  a rotationally supported disk with a size of $\sim10$\,au at the very early accretion phase (Fig.~\ref{fig:1}{\it a}).
At the same epoch, the outflow extends to $\sim20$\,au (Figs.~\ref{fig:1}{\it b} and {\it c}). 
The size growth of the rotationally supported disk can  be confirmed in the density distribution  on the $y=0$ plane (nearly edge-on view) in Figures~\ref{fig:1}{\it b}, {\it e}, {\it h} and {\it k}, in which the disk-like structure gradually increases in size. 
Although the direction of the disk normal changes over time \citep{hirano19},  the rotation motion in the disk is roughly confirmed on the $z=0$ plane (Figs.~\ref{fig:1}{\it a}, {\it d}, {\it g}, {\it j}).

Figure~\ref{fig:1} also shows that  strong mass ejection frequently occurs as the disk grows. 
Shell-like structures seen in Figures~\ref{fig:1}{\it f}, {\it h} and {\it i} are caused by episodic mass ejection \citep{hirano19}, in which each shell roughly corresponds to each episode of mass ejection. 
Although the direction of the outflow changes with time, the strong mass ejection lasts until the end of the simulation, as also seen in \citet{machida20b}.  
The maximum speed of the outflowing gas exceeds $\sim100\,\kms$. 
Violent mass ejection by the high-velocity jet can be seen in Figure~\ref{fig:1}{\it h}, {\it i}, {\it k} and {\it l}.
%%In Paper III, low-velocity outflow appears for the same model, DM3.
%%Thus, for this model, although a high-velocity jet appears, the low-velocity (wide-angle) outflow can be driven without the help of the high-velocity component. 

%%%%%%
% Fig. 2
%%%%%%
\begin{figure*}
\begin{center}
\includegraphics[width=0.8\columnwidth]{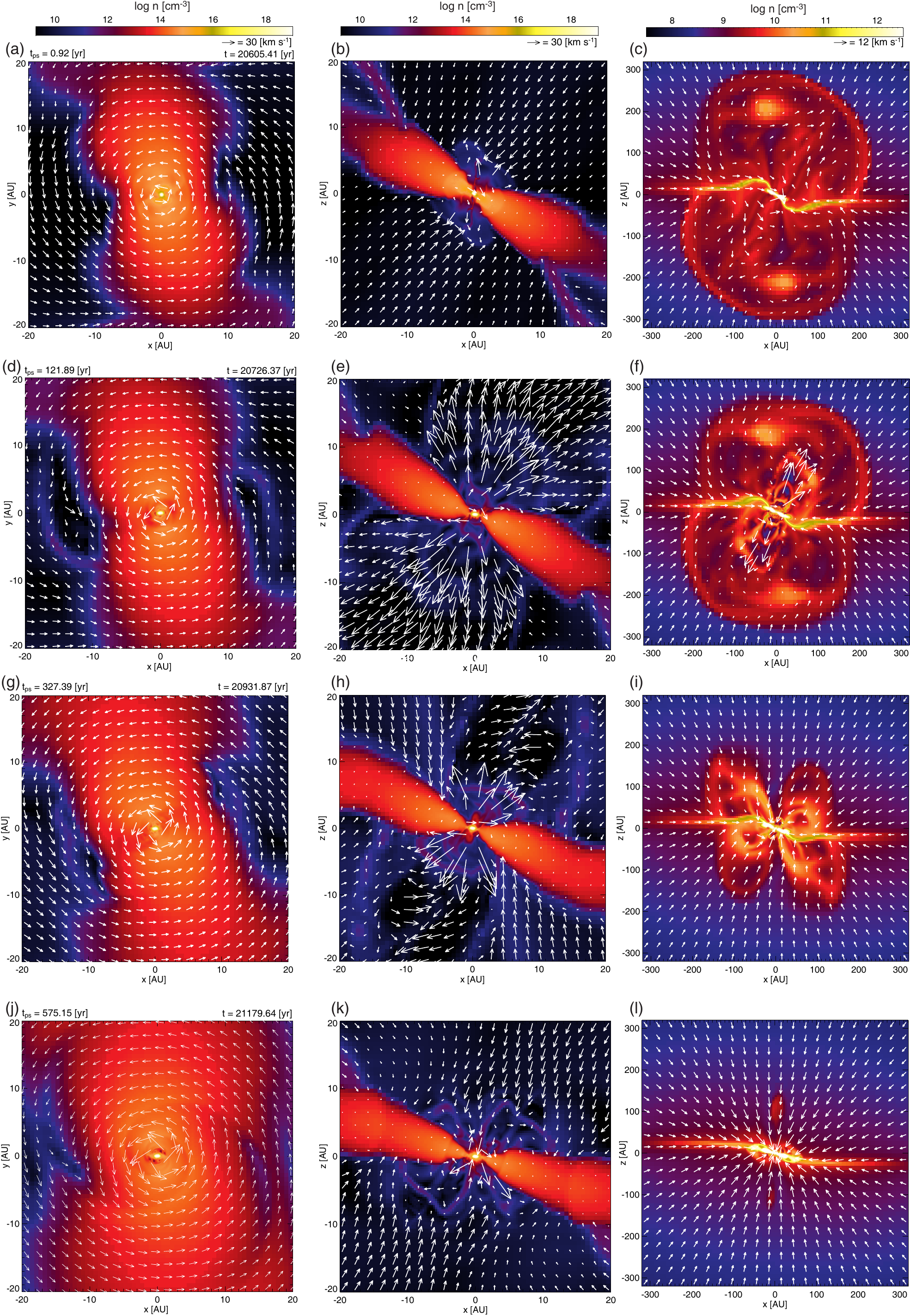}
\end{center}
\caption{
Density (color) and velocity (arrows) distributions at different epochs on the $z=0$ (left panels) and $y=0$ (middle and right panels) planes for EM3 (see also EM3.avi).  
Note that the box scale of the left and middle panels is different from that of the right panels. 
The elapsed time $t_{\rm ps}$ after protostar formation and that $t$ after the beginning of the cloud collapse are described in each left panel. 
}
\label{fig:2}
\end{figure*}

%%%%%%
% Fig. 3
%%%%%%
\begin{figure*}
\begin{center}
\includegraphics[width=0.8\columnwidth]{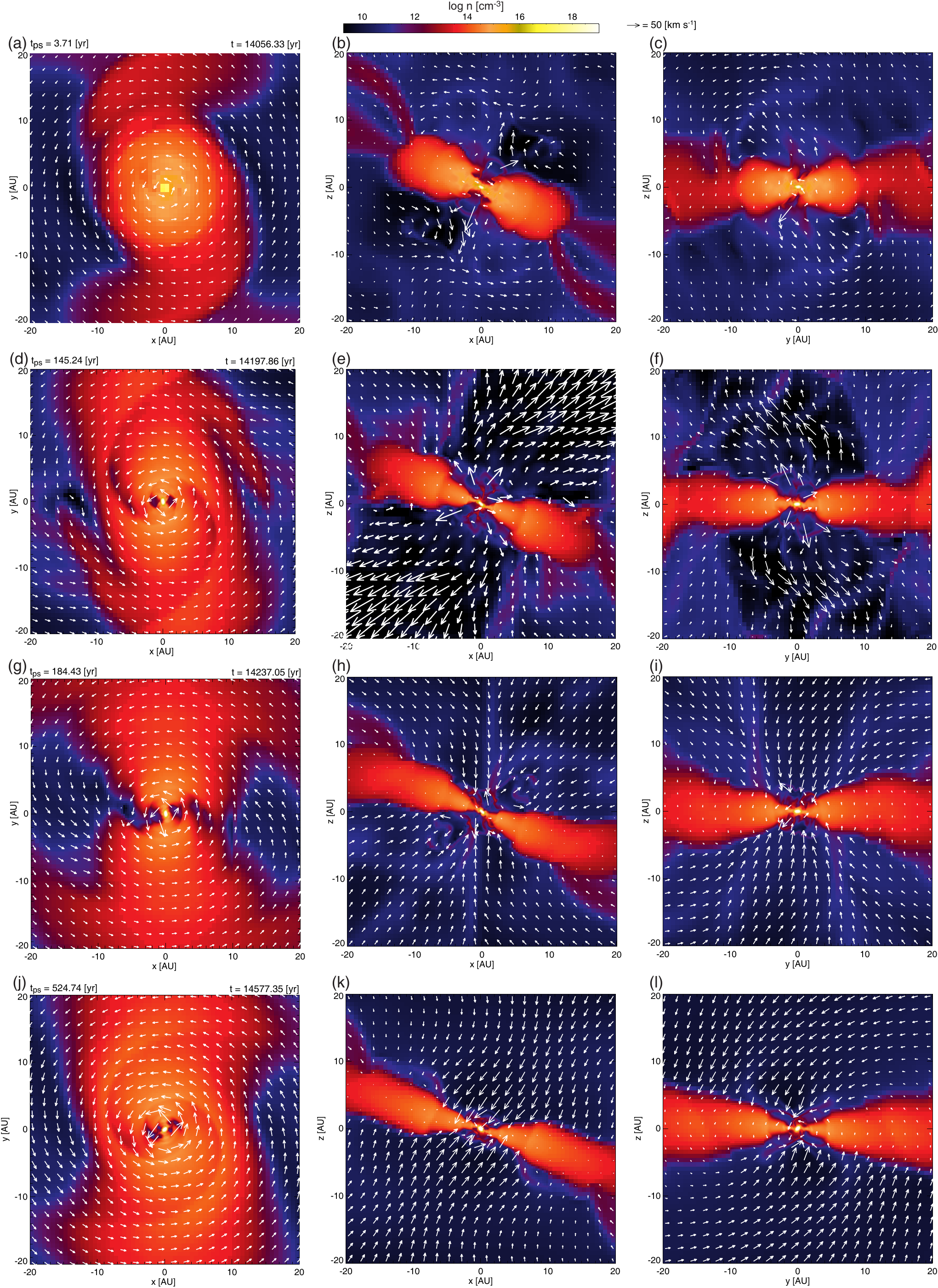}
\end{center}
\caption{
As Fig.~\ref{fig:1} but for FM3 (see also FM3.avi). 
}
\label{fig:3}
\end{figure*}
%%%%%%
% Fig. 4
%%%%%%
\begin{figure*}
\begin{center}
\includegraphics[width=0.8\columnwidth]{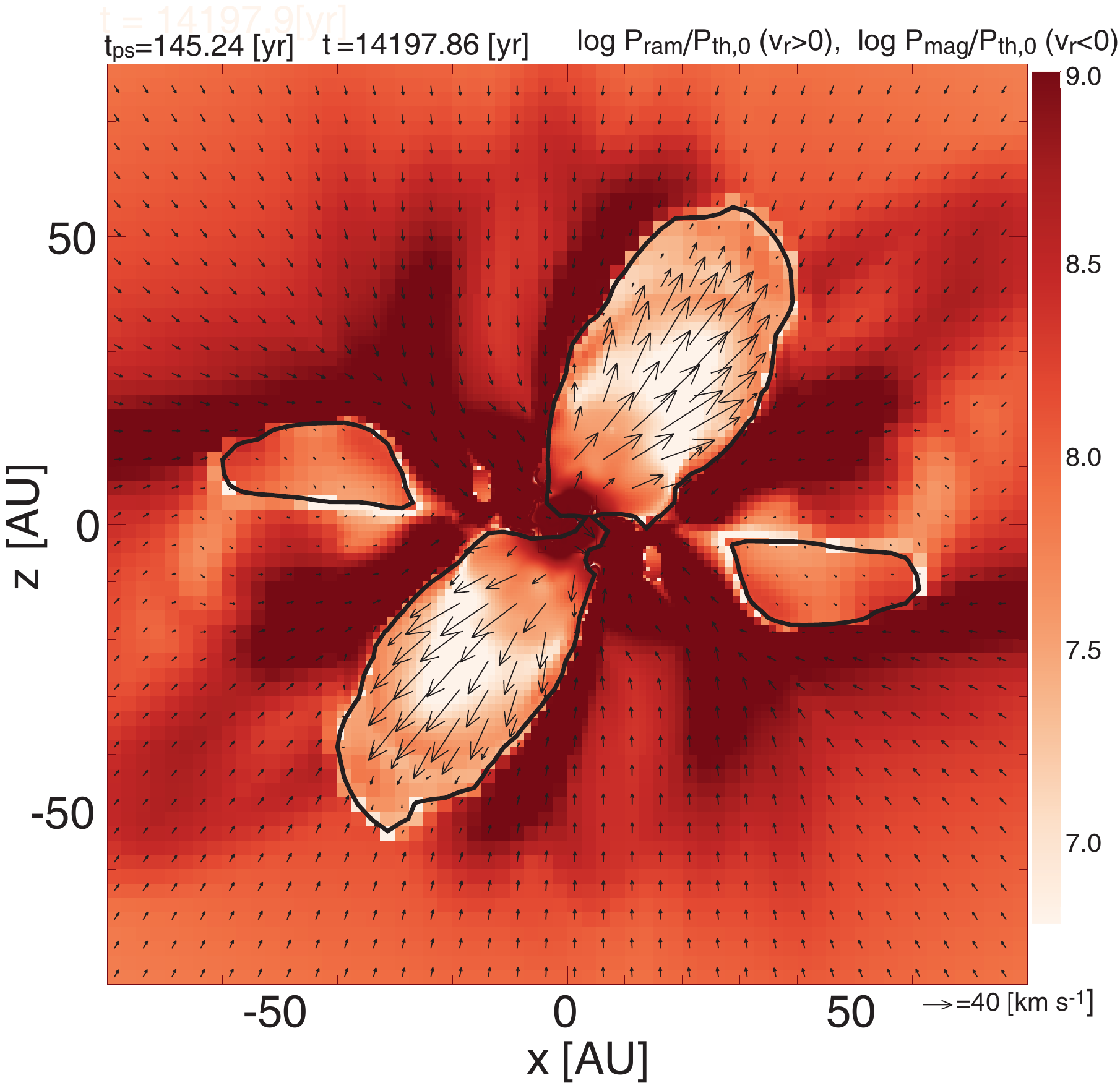}
\end{center}
\caption{
Ram and magnetic pressures on the $y=0$ plane for FM3.  
The ram pressure is plotted outside the outflow and the magnetic pressure is plotted inside the outflow. 
Both the ram ($P_{\rm ram}/P_{\rm th, 0}$) and magnetic ($P_{\rm mag}/P_{\rm th, 0}$) pressures are normalized by the initial thermal pressure at the center of the cloud $P_{\rm th, 0}$. 
The boundary between the outflow ($v_r>0$) and infalling envelope ($v_r<0$) is plotted by the black solid curve. 
The arrows represent the velocity at each point. 
The elapsed  time after protostar formation $t_{\rm ps}$ and that after the cloud begins to collapse $t$ are described in each panel. 
}
\label{fig:4}
\end{figure*}

Figure~\ref{fig:2} shows the time sequence of the region around the protostar for EM3 with parameters $\mu_0=3$ and $\alpha_0=0.02$.
For this model, the low-velocity outflow appears before protostar formation and creates a clear cocoon-like structure, as seen in Figure~\ref{fig:2}{\it c} and {\it f}. 
A high-velocity jet begins to appear after protostar formation (Fig.~\ref{fig:2}{\it b}). 
Then, strong mass ejection occurs around the protostar at $t_{\rm ps} \sim100$\,yr (Fig.~\ref{fig:2}{\it e}), where $t_{\rm ps}$ is the elapsed time after protostar formation. 
On the other hand,  Figures~\ref{fig:2}{\it c} and {\it f} indicate that the low-velocity outflow weakens and the cocoon-like structure gradually shrinks.
Figure~\ref{fig:2}{\it f} shows that the high-velocity jet propagates into the shrinking cocoon-like structure formed by the low-velocity outflow. 
The high-velocity jet also gradually loses activity for $t_{\rm ps} \gtrsim 150$\,yr (Figs.~\ref{fig:2}{\it h} and {\it i}). 
Finally, the strong mass ejection stops and  both the low-velocity outflow and high-velocity jet disappear,  as seen in Figure~\ref{fig:2}{\it k} and {\it l}. 
The regions above and below the protostar and disk are disturbed by the outflow and have complicated structures just after protostar formation (Figs.~\ref{fig:2}{\it c} and {\it f}). 
On the other hand, the density distribution is rather smooth and no particular structure is seen in these regions at $t_{\rm ps}\simeq500$\,yr (Fig.~\ref{fig:2}{\it l}).
However, a very weak jet appears in the region very close to  the protostar (Fig.~\ref{fig:2}{\it k}). 

%%In Paper III, the low-velocity outflow for this model begins to be activated again about 2,000\,yr after protostar formation. 
%%Although I calculated the cloud evolution for $\sim500$\,yr after protostar formation, the high-velocity jet does not evolve within the calculation time. 
%%Thus, it is expected that the high-velocity jet cannot contribute to the driving the low-velocity outflow for this model. 

Figure~\ref{fig:3} shows the time sequence of the region around the protostar for FM3 which has the parameters  $\mu_0=3$ and  $\alpha_0=0.01$.
For this model, no low-velocity outflow appears before protostar formation (Papers I and III). 
A cocoon-like structure can be confirmed just after protostar formation in  DM3 (Fig.~\ref{fig:1}{\it c}) and EM3 (Fig.~\ref{fig:2}{\it c}), while no such structure is seen in FM3 (Fig.~\ref{fig:3}{\it c}).
%% consistent with Paper III. 
After protostar formation, a high-velocity jet also begins to appear for this model, as seen in Figures~\ref{fig:3}{\it e} and {\it f}. 
However,  the jet rapidly weakens and disappears at $t_{\rm ps} \sim 160$\,yr (Fig.~\ref{fig:3}{\it h}). 
Then, the jet sometimes appears near the protostar, but does not grow sufficiently by the end of the simulation. 
On the other hand, the disk-like structure grows with time (see the middle panels of Fig.~\ref{fig:2}).
At the end of the simulation, we cannot see any sign of mass ejection (Figs.~\ref{fig:3}{\it k} and {\it l}).

For FM3, the ram (outside the jet) and magnetic (inside the jet) pressures when the jet is decaying (or shrinking) are plotted in Figure~\ref{fig:4}. 
The figure indicates that the ram pressure in the envelope (or outside the jet) is larger than the magnetic pressure inside the jet.
In addition, the jet velocity rapidly decreases near the boundary between the jet and infalling envelope. 
Thus, it is expected that a  high ram pressure suppresses the growth of the jet at this epoch, which is also seen in the case of decaying low-velocity outflow (Paper I and III). 
%%%%%%
% Fig. 5
%%%%%%
\begin{figure*}
\begin{center}
\includegraphics[width=0.9\columnwidth]{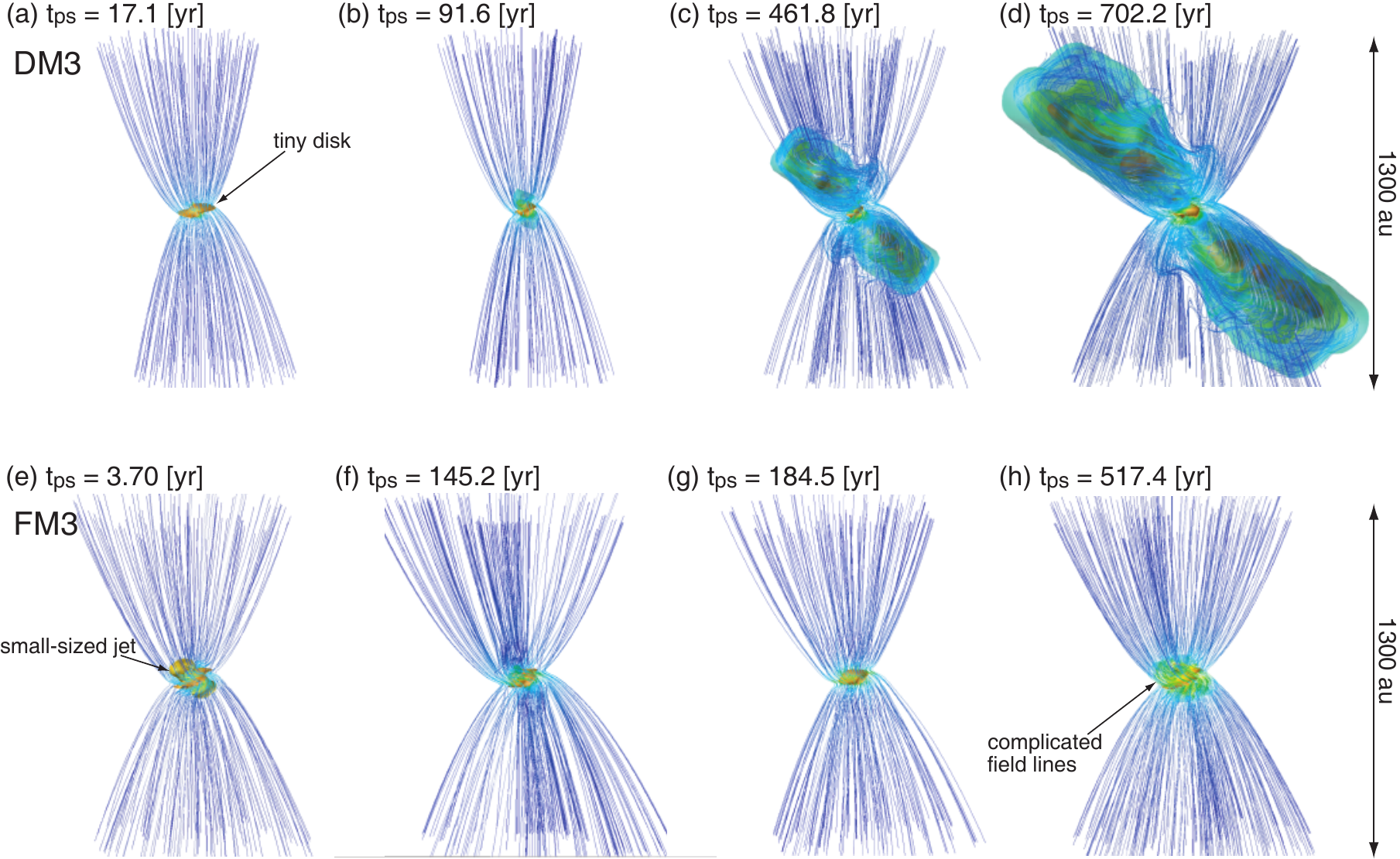}
\end{center}
\caption{
Three dimensional view of DM3 (upper panels) and FM3 (lower panels) at four different epochs. 
In each panel, the blue lines are magnetic field lines and the central orange region corresponds to the high density region (iso-density surface of $10^{12}\cc$). 
The outflowing region with $v_r>0$ is plotted by the nested color surfaces.
}
\label{fig:5}
\end{figure*}

Figure~\ref{fig:5} shows the time sequence of the structure in a three-dimensional view for DM3 (upper panels) and FM3 (lower panels).
It should be noted that since the time sequence for EM3 is essentially the same as FM3, EM3 is omitted from Figure~\ref{fig:5}.
In DM3,  we can confirm a tiny disk-like structure at the center and hourglass shaped magnetic field lines in the whole region at the early mass accretion stages  (Figs.~\ref{fig:5}{\it a} and {\it b}).  
For this model, the outflow  gradually grows with time and the magnetic field lines around the protostar are disturbed by the outflow  (Figs.~\ref{fig:5}{\it c} and {\it d}).  
In Figure~\ref{fig:5}{\it d}, the high-velocity component (jet) colored yellow and orange is enclosed by the low-velocity component (outflow) colored blue. 
The high-velocity jet, which is embedded in the low-velocity component, has a knotty structure that is caused by the episodic mass ejection. 
At the end of the simulation, the outflow structure has a size of $>1000$\,au.  

On the other hand, in FM3 (Fig.~\ref{fig:5} lower panels),  although an hourglass configuration of the magnetic field lines can be confirmed, no large scale outflow (either low-velocity outflow or high-velocity jet) appears.  
In Figure~\ref{fig:5}{\it h}, the magnetic field lines around the protostar are complicated, while  we cannot confirm the structure of  (large scale) outflow. 
Figure~\ref{fig:5} shows the apparent structural difference between DM3 (upper panels) and FM3 (lower panels).

%%%%%%
% Fig. 6
%%%%%%
\begin{figure*}
\begin{center}
\includegraphics[width=0.9\columnwidth]{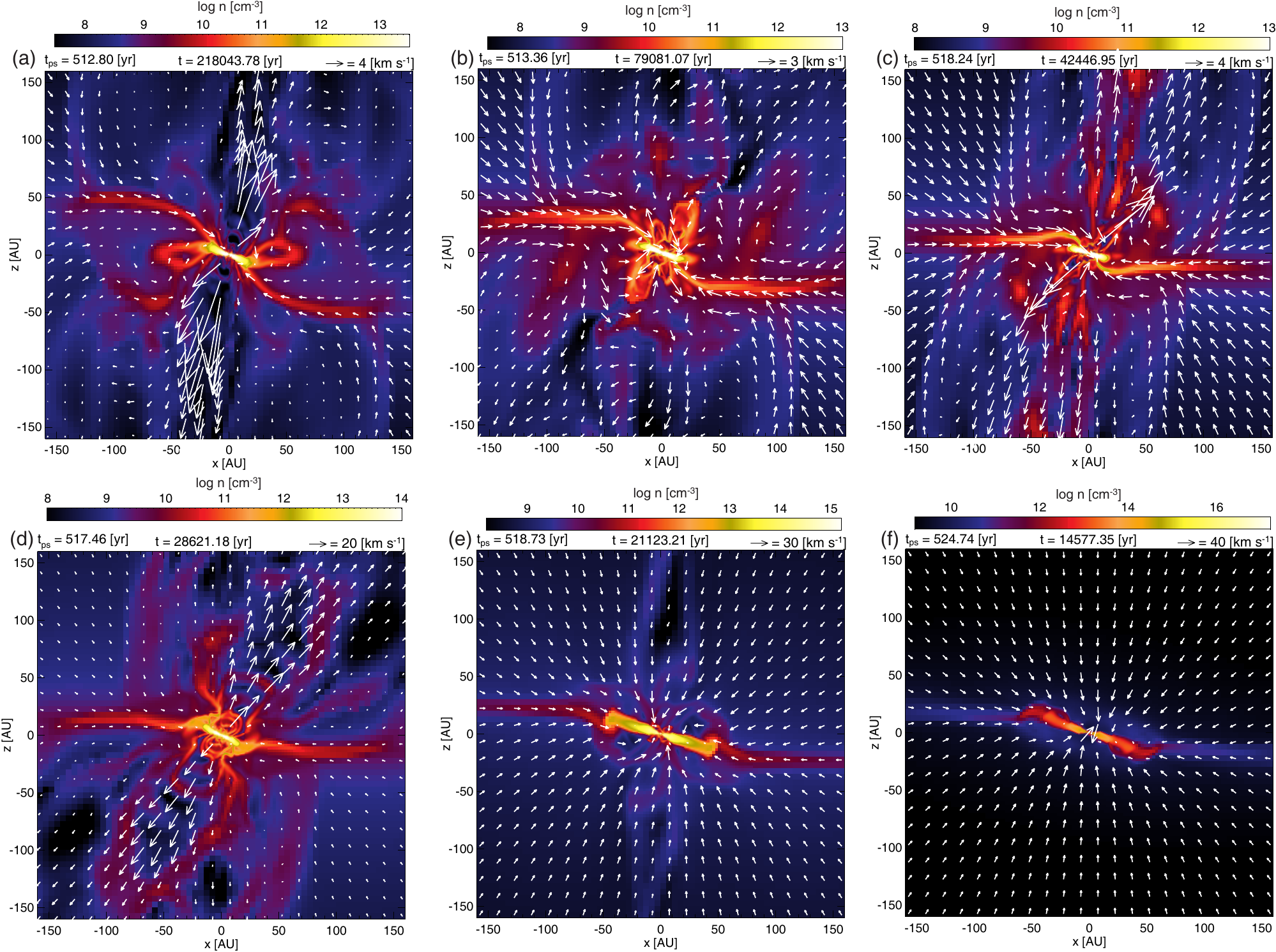}
\end{center}
\caption{
Density (color) and velocity (arrows) distributions on the $y=0$ plane at $t_{\rm ps}\sim 500$\,yr for AM3 ({\it a}), BM3 ({\it b}), CM3 ({\it c}), DM3 ({\it d}), EM3 ({\it  e}) and FM3 ({\it f}).
The elapsed time $t_{\rm ps}$ after protostar formation and that $t$ after the beginning of cloud collapse  are described in the upper part of each panel.
}
\label{fig:6}
\end{figure*}

Figure~\ref{fig:6} shows snapshots at $t_{\rm ps}\sim500$\,yr after protostar formation for all the models having $\mu_0=3$. 
In the figure, the high-velocity jet and low-velocity outflow are sustained for at least $\gtrsim 500$\,yr in AM3, BM3, CM3 and DM3, while no mass ejection occurs in EM3 and FM3 at the same epoch. 
In AM3, BM3, CM3 and DM3, although the high-velocity jet extends roughly from northeast to southwest, there is a slight difference in directions between the low-velocity outflow and the high-velocity jet. 
Especially, the jet axis is somewhat inclined from the outflow axis in CM3 (Fig.~\ref{fig:3}{\it c}), which is also seen in a recent observation \citep{matsushita19}.
The misalignment between the high- and low-velocity outflows is caused by the warp of the disk \citep[for details, see][]{hirano19,machida20b}.
For EM3 (Fig.~\ref{fig:6}{\it e}) and FM3 (Fig.~\ref{fig:6}{\it f}), the disk normal is also somewhat inclined from the $z$-axis because the initial rotation axis is set to be inclined  from the $z$-axis toward the $x$-axis. 
Figure~\ref{fig:6} indicates that the structure around the protostar differs considerably depending on the parameter $\alpha_0$ or the mass accretion rate (for details, see below). 

%%%%%%
% Fig. 7
%%%%%%
\begin{figure*}
\begin{center}
\includegraphics[width=1.0\columnwidth]{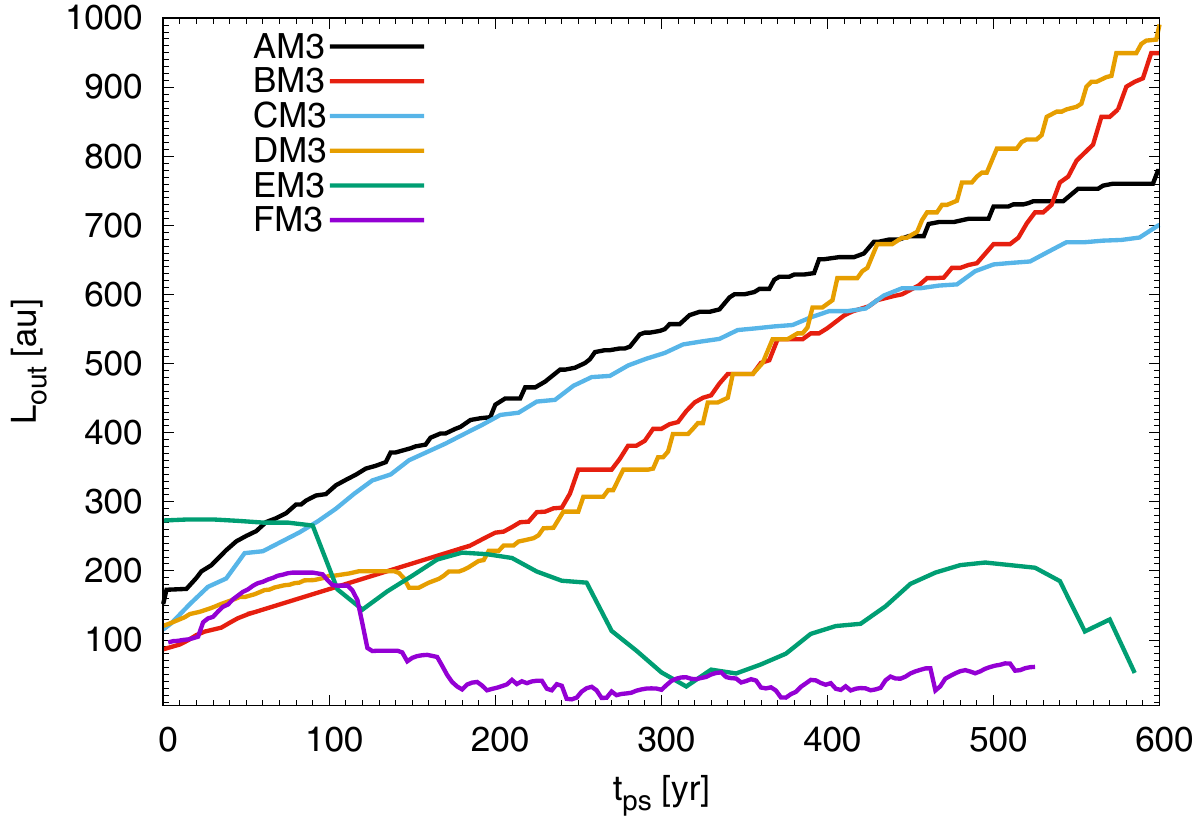}
\end{center}
\caption{
Outflow length  against the elapsed time after protostar formation for AM3, BM3, CM3, DM3, EM3 and FM3. 
}
\label{fig:7}
\end{figure*}

%%%%%%
% Fig. 8
%%%%%%
\begin{figure*}
\begin{center}
\includegraphics[width=1.0\columnwidth]{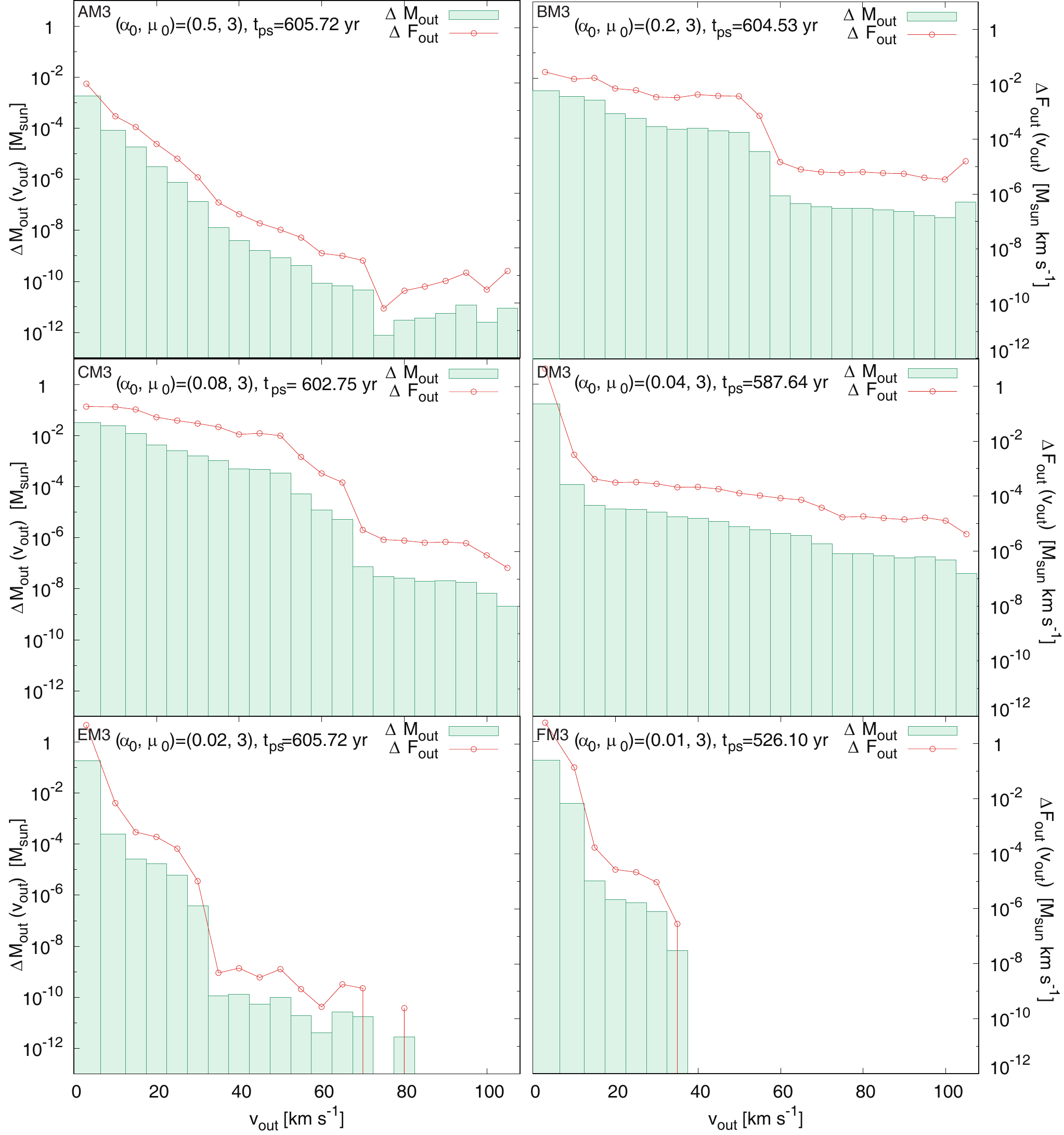}
\end{center}
\caption{
Histogram of outflow mass $\Delta M_{\rm out}$ (left axis) and momentum flux $\Delta F_{\rm out}$ (right axis) against the outflow velocity  $v_{\rm out}$ for AM3, BM3, CM3, DM3, EM3 and FM3. 
The model name, parameters ($\alpha_0$  and $\mu_0$) and elapsed time ($t_{\rm ps}$) are described in each panel.  
}
\label{fig:8}
\end{figure*}

%%%%%%
% Fig. 9
%%%%%%
\begin{figure*}
\begin{center}
\includegraphics[width=1.0\columnwidth]{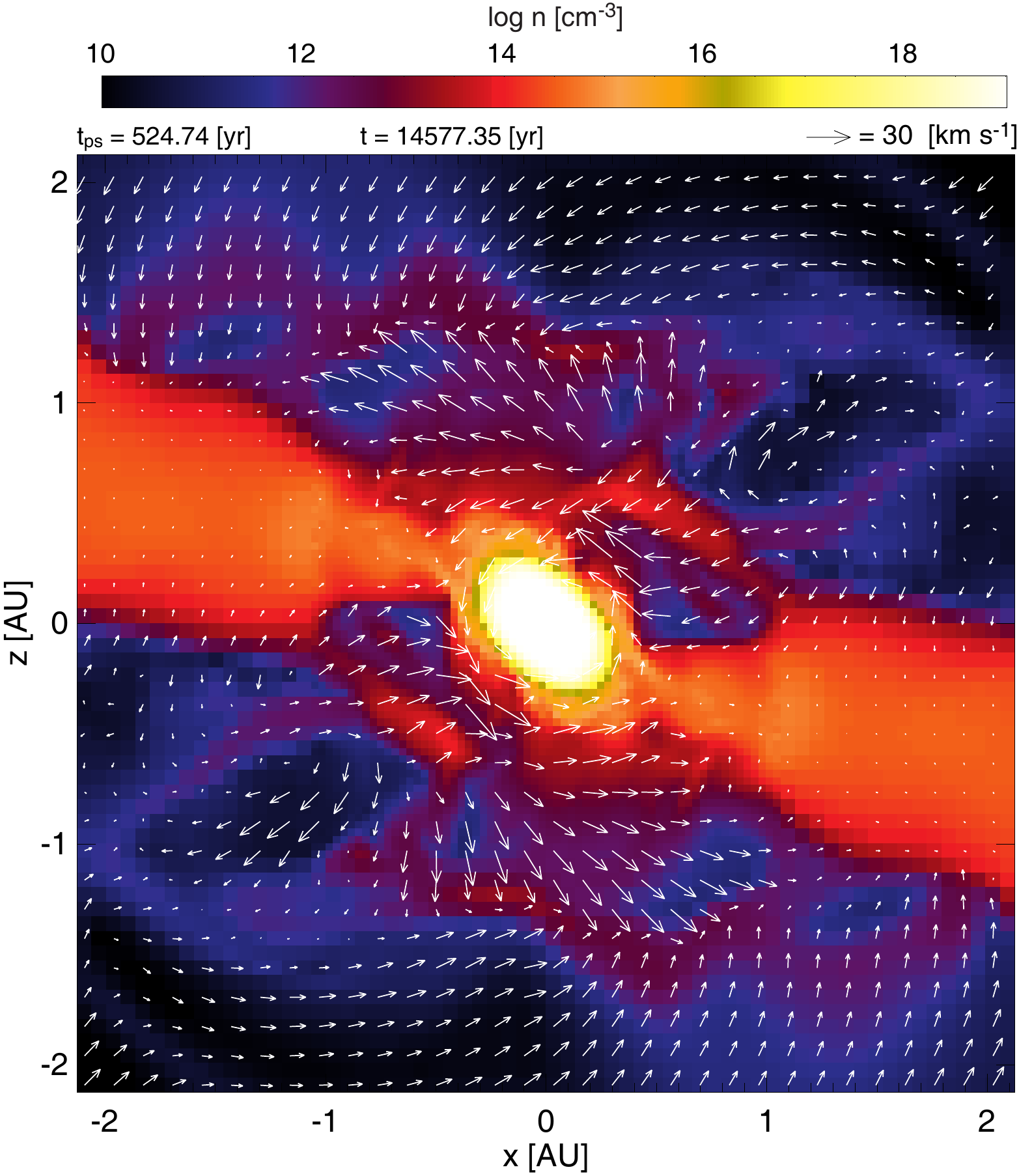}
\end{center}
\caption{
Density (color) and velocity (arrows) distributions on the $y=0$ plane for FM3 at the same epoch as in Fig.~\ref{fig:3}{\it k}.
The elapsed time $t_{\rm ps}$ after protostar formation and that $t$ after the beginning of the cloud collapse are described. 
}
\label{fig:9}
\end{figure*}

The time evolution of the outflow length for the models having $\mu_0=3$ is plotted in Figure~\ref{fig:7}, in which the outflowing gas is identified with $v_r>v_{\rm thr}$  and $v_{\rm thr}=1\,\kms$ is adopted.
I confirmed that the outflow physical quantities do not significantly depend on $v_{\rm thr}$ in the range of $0< v_{\rm thr} <3 \,\kms$.  
The figure indicates that the size of the outflow increases with time and reaches $\sim700$--$1000$\,au for about 600\,yr after protostar formation in AM3, BM3, CM3 and DM3. 
On the other hand, in EM3 and FM3, the outflow does not grow after protostar formation. 
In these models, the outflow length repeatedly increases and decreases within the range of $\lesssim 200$\,au, indicating that the high-velocity jet and low-velocity outflow appearing episodically around the protostar lose their power before the outflow grows sufficiently. 
The suppression of the outflow due to the ram pressure of the infalling envelope was elaborately investigated in Paper III (see also Fig.~\ref{fig:4}). 

To confirm the velocity component of the outflow, we plot the mass $\Delta M_{\rm out} (v_{\rm out})$ and momentum flux $\Delta F_{\rm out} (v_{\rm out})$ of the outflowing gas at each velocity component (or velocity bin), estimated as 
\begin{equation}
\Delta M_{\rm out}(v_{\rm out}) = \int_{v_{\rm out}-\Delta v}^{v_{\rm out}+\Delta v} \rho \, dV,
\label{eq:outflowmass}
\end{equation}
\begin{equation}
\Delta F_{\rm out}(v_{\rm out}) = \frac{\Delta P(v_{\rm out})}{ t_{\rm ps}},
\label{eq:momentumflux}
\end{equation}
where $\Delta v=2.5\,\kms$ was adopted, against the outflow velocity  at $t_{\rm ps} \simeq 500-600$\,yr in Figure~\ref{fig:8}.
In equation~(\ref{eq:momentumflux}), the outflow momentum $\Delta P_{\rm out}(v_{\rm out})$ at each velocity component is calculated as
\begin{equation}
\Delta P_{\rm out}(v_{\rm out}) = \int_{v_{\rm out}-\Delta v}^{v_{\rm out}+\Delta v} \rho\, \vert \vect{v} \vert \, dV.
\label{eq:momentum}
\end{equation}
Note that, in equations~(\ref{eq:outflowmass})--(\ref{eq:momentum}), the quantities are integrated only in the outflowing region of $v_r>v_{\rm thr}$. 
In AM3, BM3, CM3 and DM3, although the low-velocity components dominate the high-velocity components, there exists a non-negligible amount of outflowing mass and momentum flux in the high-velocity components. 
For these models, although the distributions of $\Delta M_{\rm out}$ and $\Delta F_{\rm out}$ show time variability, neither the low- or high-velocity components disappear at any epoch.  
A similar distribution to these models can be confirmed in the low-mass star formation case \citep{saiki20}.

On the other hand, no high-velocity component exists in EM3 and FM3 in Figure~\ref{fig:8}. 
In EM3, the mass and momentum flux of the outflowing gas rapidly decrease around $v_{\rm out}\simeq 30\,\kms$ and disappear in the range $v_{\rm out} >80\,\kms$.
In FM3, there is no high-velocity component for $v_{\rm out}> 40\,\kms$. 
Thus, it is clear that no high-velocity jets appear in EM3 and FM3. 
In these models, it seems that there are low-velocity outflow components.
However, in fact, the low-velocity components can be identified only around the region very close to the protostar and there is no evolved low-velocity outflow by the end of the simulation, as seen in Figures~\ref{fig:2} and  \ref{fig:3}.
To confirm the region around the protostar, a close-up view of Figure~\ref{fig:3}{\it k} is plotted in Figure~\ref{fig:9}. 
We cannot clearly identify the outflow in a large scale (Figs.~\ref{fig:3}{\it k} and \ref{fig:6}{\it f}), while we can confirm the outflow around the protostar in a small scale (Fig.~\ref{fig:9}).  
A major component of the outflow stays within $\sim$2\,au and does not grow by the end of the simulation for  FM3 (and EM3).

%%%%%%
% Fig. 10
%%%%%%
\begin{figure*}
\begin{center}
\includegraphics[width=0.7\columnwidth]{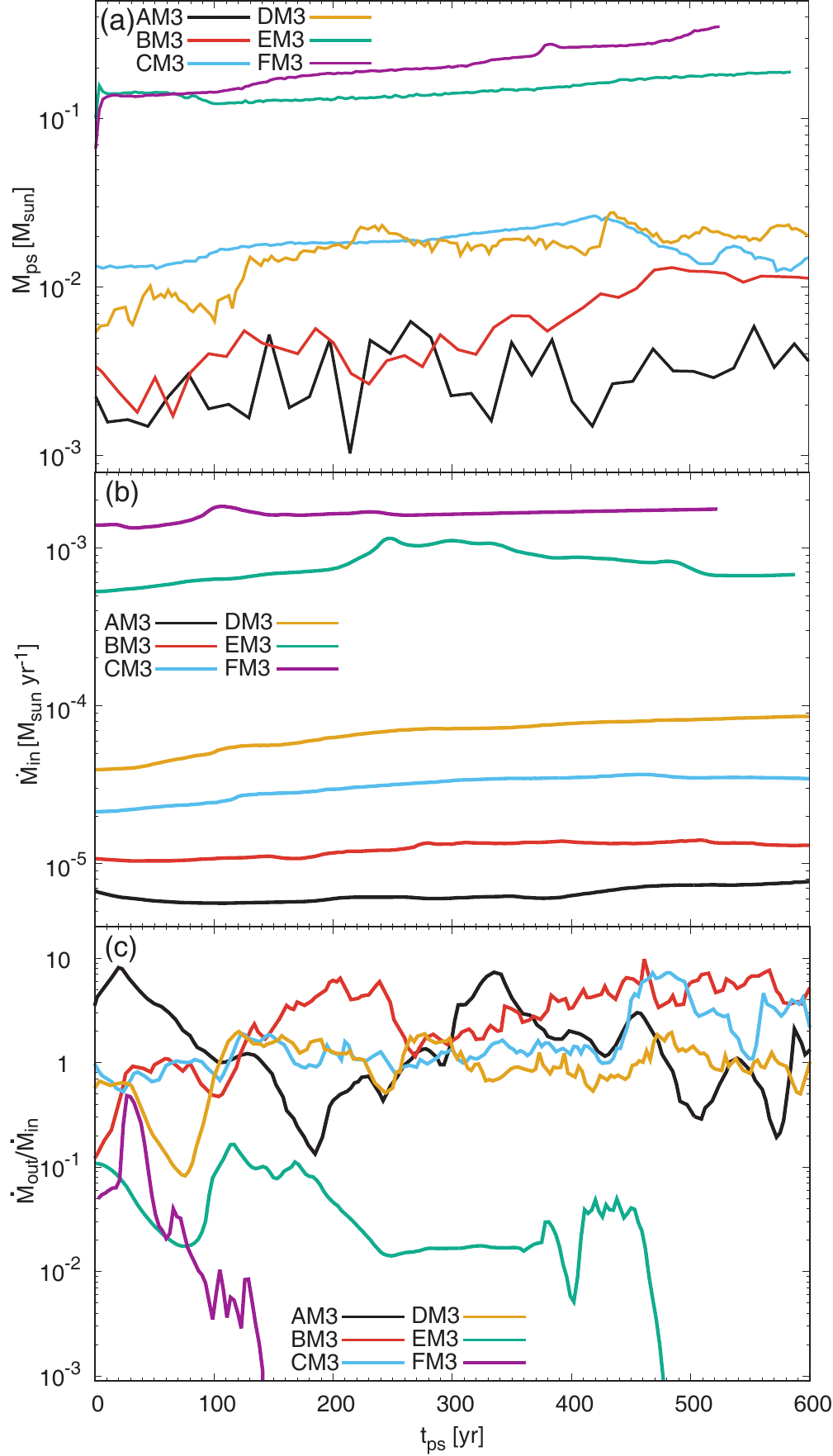}
\end{center}
\caption{
Protostellar mass ({\it a}),  inflow rate ({\it b})  and ratio of outflow to inflow rate ({\it c}) against the elapsed time after protostar formation for AM3--FM3. 
}
\label{fig:10}
\end{figure*}

Figure~\ref{fig:10} shows the protostellar mass (Fig.~\ref{fig:10}{\it a}), the mass infall rate (Fig.~\ref{fig:10}{\it b}) and the ratio of outflow to inflow rate  (Fig.~\ref{fig:10}{\it c}) against the elapsed time after protostar formation. 
The protostellar mass is estimated as being the region with $n > n_{\rm ps}$, where $n_{\rm ps}=10^{18}\cc$ is adopted \citep{machida19}.  
As seen in Figure~\ref{fig:10}{\it a},  the protostellar mass increases as $\alpha_0$ decreases,  because the mass accretion rate or infall rate depends on $\alpha_0$, as explained  in Paper I. 
The infall rate is estimated by integrating over the gas with $v_r<0$ passing through a spherical surface with a radius of 1\,au  as 
\begin{equation}
\dot{M}_{\rm in} = \int^{v_r<0} \rho\, v_r \, dS.
\end{equation} 
In the analytical estimate, the mass accretion (or infall) rate is proportional to $\alpha_0^{-3/2}$ (for details, see Paper I).
Thus, it is natural that the mass accretion rate for models with small $\alpha_0$ is higher than that for the model with large $\alpha_0$. 

We can see a significant difference (or gap) between DM3 and EM3 in Figures~\ref{fig:10}{\it a} and {\it b}
caused by the lack of outflow  in EM3 and FM3. 
Figure~\ref{fig:10}{\it c} shows the ratio of outflow to inflow rate, in which the outflow rate, 
\begin{equation}
\dot{M}_{\rm out}= \int^{v_r>0} \rho\, v_r \, dS,   
\end{equation}  
is estimated using the same spherical surface as used in estimating the inflow rate ($\dot{M}_{\rm in}$).
In the panel, we can also see a significant difference between the models with and without outflow. 
As shown in Figures~\ref{fig:6} and \ref{fig:7}, the outflow grows in AM3, BM3, CM3 and DM3, while it is not evolved in EM3 and FM3.

For AM3, BM3, CM3 and DM3, the outflow rate is comparable or larger than the infall rate,
as seen in Paper I. 
The infall and outflow rates are estimated at a radius of $1$\,au and the protostar has a size of 0.01--0.2\,au.
Thus,  a part of the infalling gas is swept by the high-velocity jet  driven near the protostar. 
As a result, the mass of the outflowing gas can dominate that  of the infalling gas (see also Papers I and II).
Since not all the infalling gas is ejected by the high-velocity jet, the protostellar mass gradually increases even when strong mass ejection occurs around the protostar (AM3, BM3, CM3 and DM3). 

On the other hand, the ratio $(\dot{M}_{\rm out}/\dot{M}_{\rm in})$ rapidly decreases in EM3 and FM3 (Fig.~\ref{fig:10}{\it c}). 
Since the mass inflow rate is high in such models, the decrease of the ratio indicates that the mass outflow rate rapidly decreases. 
Thus, the infalling gas is not significantly disrupted by the high-velocity jet in these models. 
Therefore, a high infall rate (Fig.~\ref{fig:10}{\it b}) and high-mass protostar tends to be realized (Fig.~\ref{fig:10}{\it c}) in these models.  
Figure~\ref{fig:10}{\it a} indicates that the outflow affects the mass growth of the protostar. 

%%%%%%
% Fig. 11
%%%%%%
\begin{figure*}
\begin{center}
\includegraphics[width=0.9\columnwidth]{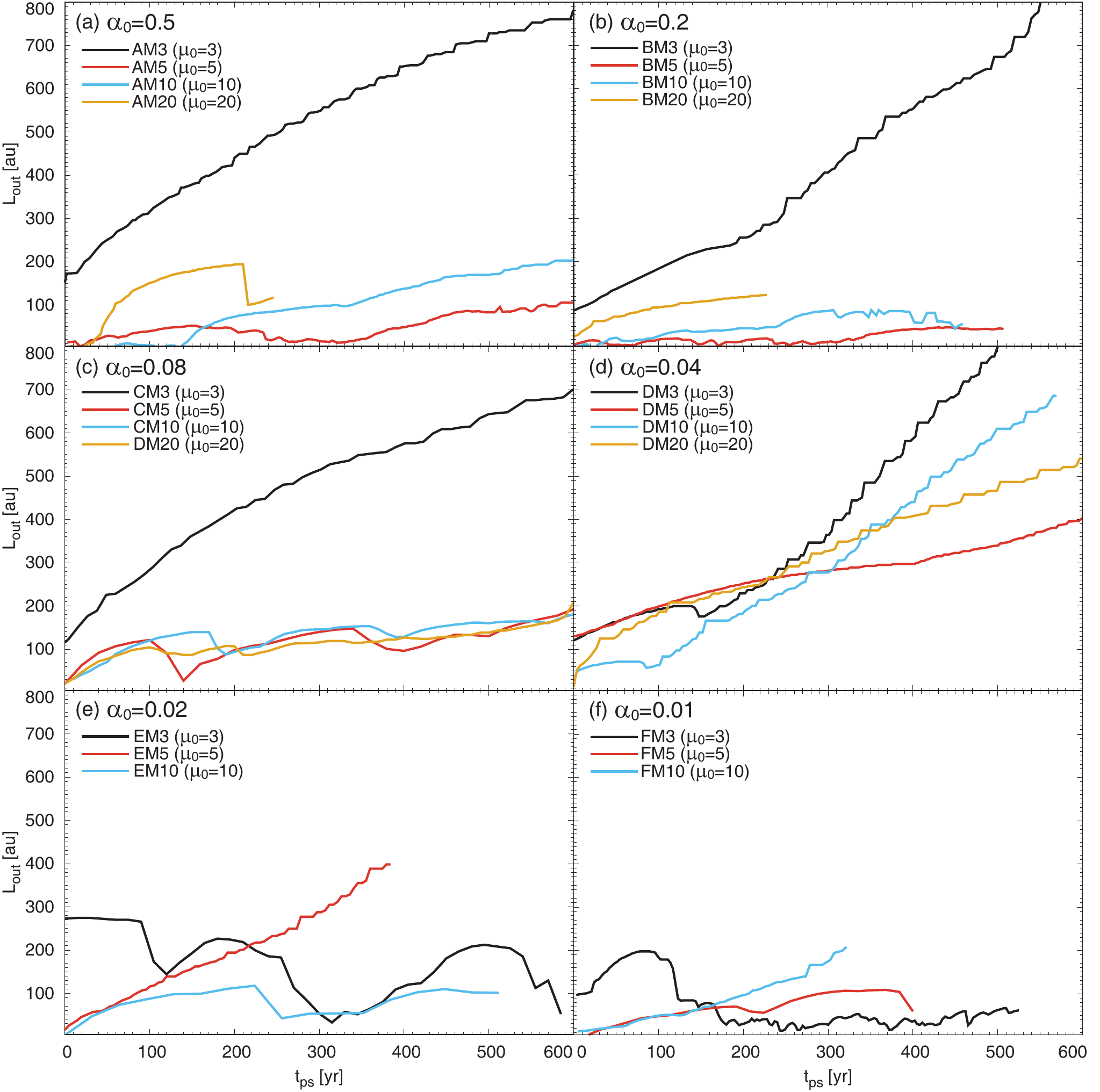}
\end{center}
\caption{
Outflow length against the elapsed time after protostar formation for all models.
The model name and parameters $\alpha_0$ and $\mu_0$ are described in each panel. 
}
\label{fig:11}
\end{figure*}

\subsection{Parameter Dependence}
While we considered the cloud evolution for the models with an initially strong magnetic field with $\mu_0=3$ in \S\ref{sec:strong}, 
in this subsection, the cloud evolution for all models is described.
Figure~\ref{fig:11} shows the outflow length  against the elapsed time after protostar formation for all the models listed in Table~\ref{table:1}. 
In the figure, for the models with $\alpha_0=0.5$ (Fig.~\ref{fig:11}{\it a}), 0.2 (Fig.~\ref{fig:11}{\it b}) and 0.08 (Fig.~\ref{fig:11}{\it c}), the outflow length continues to increase only when the initial magnetic field is as strong as  $\mu_0=3$. 
Figures~\ref{fig:11}{\it a}--{\it c} also show that  the outflow lengths for the models with $\mu_0\ge 5$ stay within $\sim200$\,au by the end of the simulation
\footnote{
A sharp drop of AM20 at $t_{\rm ps}\sim200$\,yr seen in Fig.~\ref{fig:11}{\it a} is due to the outflow identification criteria. 
As described in \S\ref{sec:strong}, I defined the outflow as the outflowing gas having $v_r\ge1\,\kms$.
The head of the outflow gradually decelerates, while the mass ejection episodically occurs.
Thus,  the inner outflowing gas is identified as the head of the outflow after the forward outflow decelerates  to $v_r<1\,\kms$, which causes the sharp drop of the outflow length. 
}. 
Thus, neither the low-velocity outflow nor high-velocity jet sufficiently evolve in these models. 
These models ($\alpha_0=0.5$, 0.2 and 0.08) have a relatively low mass accretion rate of $\dot{M}\lesssim 5\times 10^{-5}\msun$\,yr$^{-1}$ (Fig.~\ref{fig:10}{\it b}).

The models with $\alpha_0=0.04$ (Fig.~\ref{fig:11}{\it d}) are an exception. 
As seen in Figure~\ref{fig:10}{\it b}, the model with $\alpha_0=0.04$, DM3,  has an intermediate mass accretion rate of $\sim 10^{-4}\,\msun$\,yr$^{-1}$.
In Figure~\ref{fig:11}{\it d}, although the outflow growth rates are different among the models, the outflows continue to evolve up to the end of the simulation in all the models (DM3, MD5, DM10 and DM20).

For the models with $\alpha_0=0.02$ and 0.01 (EM3, EM5, EM10, FM3, FM5, FM10), which have a high mass accretion rate  ($> 10^{-4}\,\msun$\,yr$^{-1}$), the low-velocity outflow and high-velocity jet are not very active. 
In Figures~\ref{fig:11}{\it e} and {\it f}, the outflow lengths  for these models oscillate within 300\,au.
Thus, almost none of the models plotted in these panels show a successful growth of the outflow. 
Among the models, although EM5 and FM10 show continuous growth of the outflow, their lengths are much smaller than those of the successful growth models (AM3, BM3, CM3, DM3).

%%%%%%
% Fig. 12
%%%%%%
\begin{figure*}
\begin{center}
\includegraphics[width=0.9\columnwidth]{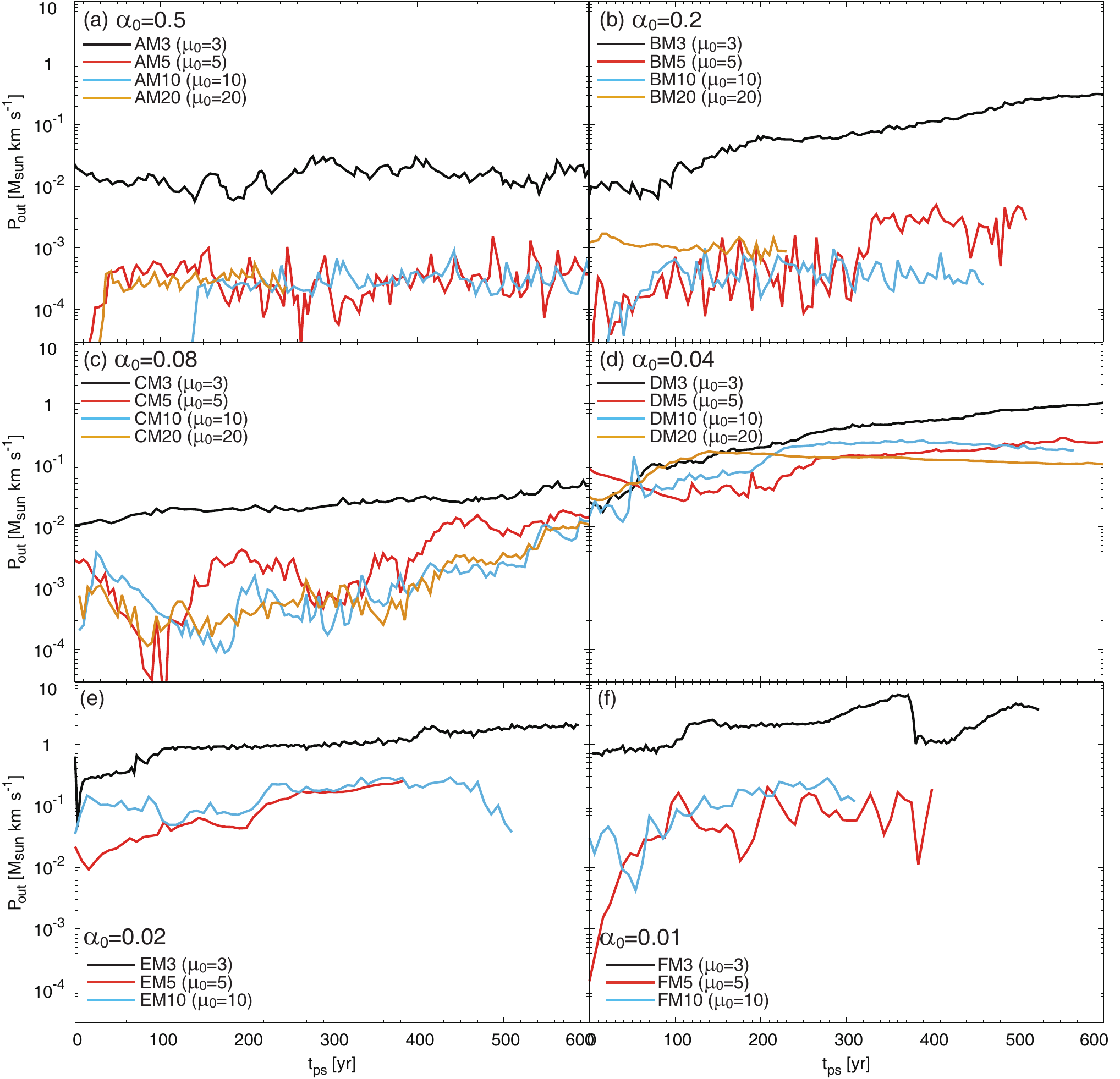}
\end{center}
\caption{
Outflow momentum for all models against the elapsed time after protostar formation.
The model name and parameters $\alpha_0$ and $\mu_0$ are described in each panel. 
}
\label{fig:12}
\end{figure*}

Figure~\ref{fig:12} shows the outflow momentum $P_{\rm out}$, calculated as 
\begin{equation}
P_{\rm out} = \int^{v_r > v_{\rm thr}} \rho \vert \vect{v} \vert dV, 
\end{equation} 
for all model.
The trend of the outflow momentum seen in Figure~\ref{fig:12} is similar to that in Figure~\ref{fig:11}.
In Figure~\ref{fig:12}{\it a}--{\it c} (models with $\alpha_0=0.5$, 0.2 and 0.08), the outflow momenta for the models with $\mu_0=3$ are much larger than those for models with $\mu_0\ge5$. 
For the models with $\alpha_0=0.04$ (DM3, DM5, DM10, DM20; Fig.~\ref{fig:12}{\it d}), the difference in the outflow momenta among the models  is not very large, because the outflow in these models continues to grow up to the end of the simulation, as shown in Figure~\ref{fig:11}.
However, the outflow momentum is larger in the model with a stronger magnetic field than that with a weaker magnetic field at the end of the simulation.

For the models with high mass accretion ($\alpha_0=0.02$ and 0.01; Figs.~\ref{fig:12}{\it e} and {\it f}), the outflow momenta  for the models with the strongest magnetic field (EM3, FM3) are much larger than those with weak magnetic fields (EM5, EM10, FM5, FM10).  
However, Figures~\ref{fig:11}{\it e} and {\it f} indicate that the outflow does not sufficiently grow and the outflowing gas stays near the protostar for the models with $\mu_0=3$ (EM3, FM3). 
Thus, the large momenta for these models is due to the dense gas staying around the protostar (see Figs~\ref{fig:2}, \ref{fig:3} and \ref{fig:9}).  

In addition, Figures~\ref{fig:11}{\it e} and {\it f} also indicate that the outflow can grow in models EM5 and FM10. 
However, the outflow momenta for models EM5 and FM10 are about one order of magnitude smaller than those for the models with $\mu_0=3$ (EM3, FM3). 
Thus, it is considered that only a very weak outflow can evolve in such models. 

%%%%%%
% Fig. 13
%%%%%%
\begin{figure*}
\begin{center}
\includegraphics[width=0.9\columnwidth]{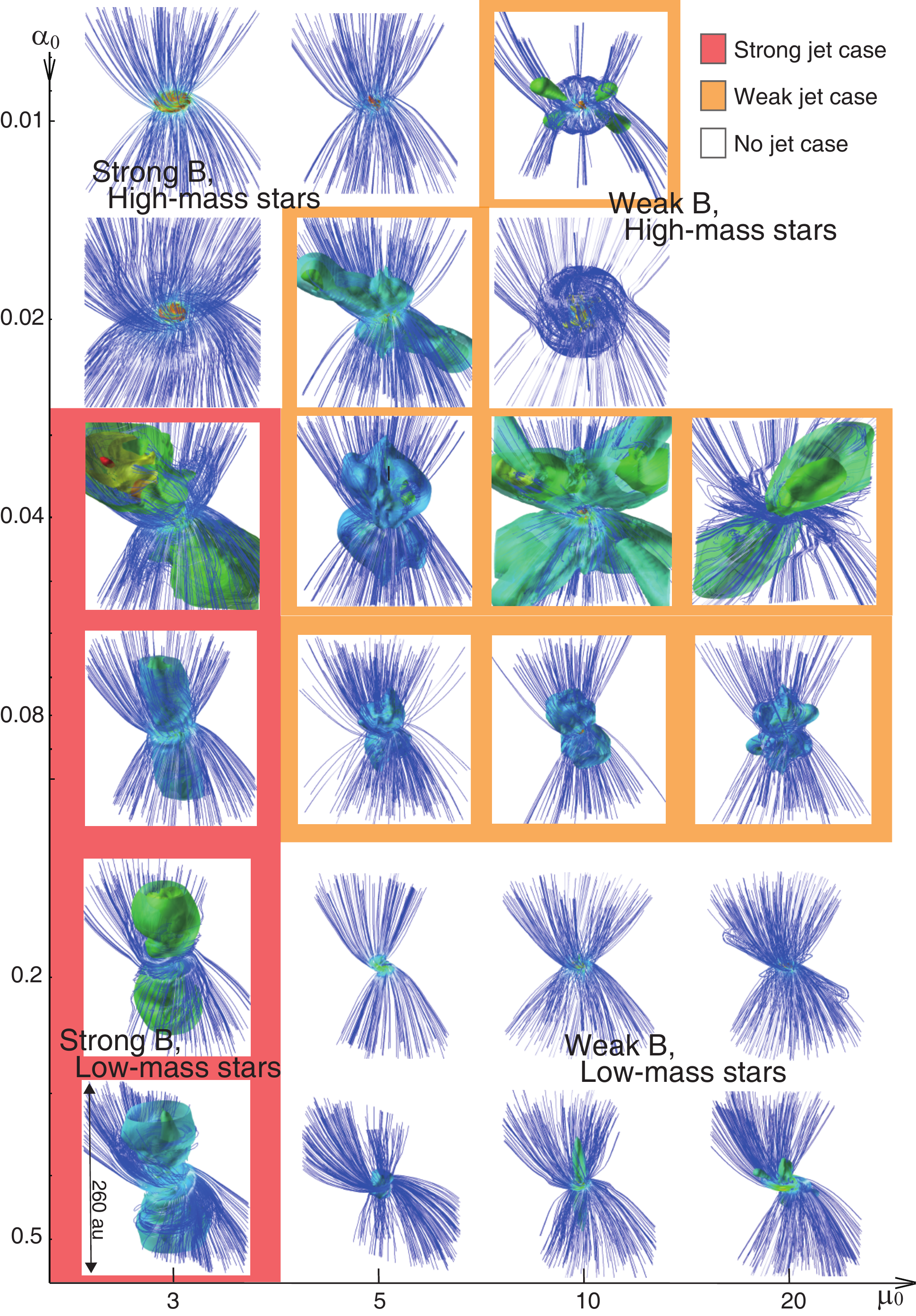}
\end{center}
\caption{
Three dimensional structure at the end of the simulation plotted on the parameter plane ($\mu_0$ and $\alpha_0$).
In each panel, the magnetic field lines (blue stream lines), outflowing component (green, yellow and red iso-velocity surfaces) and disk-like structure (orange iso-density surface) are plotted. 
Strong (red), weak (orange) and no (white) jet models are represented by the background color. 
}
\label{fig:13}
\end{figure*}

Figure~\ref{fig:13} shows the three dimensional structure at the end of the simulation for all the models  plotted on the $\mu_0$--$\alpha_0$ parameter plane. 
The box size  (260\,au) is the same in each panel.
As described in \S\ref{sec:settings}, the parameter $\mu_0$ corresponds to the magnetic field strength of the prestellar cloud, while the parameter $\alpha_0$ controls the mass accretion rate onto the central region. 
In the figure, the models are classified into three categories as strong, weak  and no  jet cases (or models), with red, orange and white backgrounds, respectively. 
Although it is difficult to determine the jet strength (strong or weak jet), I used the momentum (Fig.~\ref{fig:12}) and length growth (Fig.~\ref{fig:11}) of the outflowing gas  as indicators of the jet strength. 
Note that the physical quantities of the outflowing gas such as outflow mass, momentum and kinetic energy increase as the mass accretion rate increases or the parameter $\alpha_0$ decreases (for details, see Papers I and II). 
Thus,  I only compared the outflow physical quantities (Figs.~\ref{fig:11} and \ref{fig:12}) for the models with the same $\alpha_0$. 

In Figure~\ref{fig:13}, the strong jet models (red background) are distributed only in the lower left, indicating that a strong (high-velocity) jet appears only when both conditions of strong magnetic field and low-mass accretion rate are realized. 
A clear bipolar jet structure can been seen in the models in red panels.
In the models located in the lower center and right (panels without background color), although the magnetic field lines are slightly entangled around the center, we cannot confirm a clear jet structure.   

We can confirm a jet (or outflow) structure in all the models with an intermediate accretion rate ($\alpha_0=0.04$ and 0.08 models) in Figure~\ref{fig:13}.
A clear bipolar structure can be seen in the models with $\mu_0=3$. 
On the other hand, the structures of the jet (or outflow) are highly complicated or  small for the models with $\mu_0 \ge 5$.    
For the high mass accretion cases with $\alpha_0=0.01$ and 0.02, the models with the strongest magnetic fields ($\mu_0=3$) do not show a jet structure. 
Meanwhile, a bipolar jet-like structure can be seen in the models with ($\alpha_0$, $\mu_0$) = (0.02, 5) and (0.01, 10). 

In summary, the models with a strong magnetic field and/or a low-mass accretion rate always show a clear jet structure.
The models with a high-mass accretion rate and a weak magnetic field  do not always show a bipolar jet and outflow structure. 
Even if an outflow  appears in such models, the momentum and size of the outflow are very small.

%%%%%%%%%%
\section{Discussion}
%%%%%%%%%%
In the star forming regions, protostellar jets are observed as proof of star formation. 
Jet and outflow driving in the star formation process has been investigated in core collapse simulations. 
However, in many simulations, the jet driving was investigated only in a limited parameter range \citep{tomida13}.
Clouds with strong magnetic fields and low-mass accretion rates are usually chosen as the initial condition of core collapse simulations 
to investigate a  high-velocity flow driven  by the protostar
\citep{tomida13,machida14,machida19}\footnote{
It should be noted that, for the low-mass star formation case,  \citet{bate14} investigated the jet driving in the clouds with different magnetic field strengths ($\mu_0=5$, 10, 20, 100) resolving the protostar.
%% adopting initially uniform sphere, in which the protostar was spatially resolved. 
Although they calculated the jet driving only for $\sim2-3$\,yr, the high-velocity jet extends to $\sim3$\,au from the protostar in the models with $\mu_0=5$, 10, and 20. 
Note that, in their study, no jet appears in the model with $\mu_0=100$. 
I confirmed that, in my simulations, the jet appears and is sustained at least in the very early phase for the model with $\mu_0=3$, 5, 10 and 20 (Fig.~\ref{fig:11}). 
}. 
{
The difference in the initial condition may cause different outcomes \citep{vaytet18,wurster18}.
}
Thus, we need to perform core collapse simulations with various initial conditions to investigate whether protostellar  jets universally appear in the star formation process. 
This study focuses on jet driving with different star-forming environments. 
In other words, I especially included jet driving in clouds with weak magnetic fields and a high mass accretion rates in the investigation.
Another purpose of this study is to investigate whether a jet driven near the protostar can help to drive a low-velocity and wide-angle outflow, based on the finding in Paper III that the outflow fails to be driven with a weak magnetic field and/or a high mass accretion rate. 

%%%%%%
% Fig. 14
%%%%%%
\begin{figure*}
\begin{center}
\includegraphics[width=0.9\columnwidth]{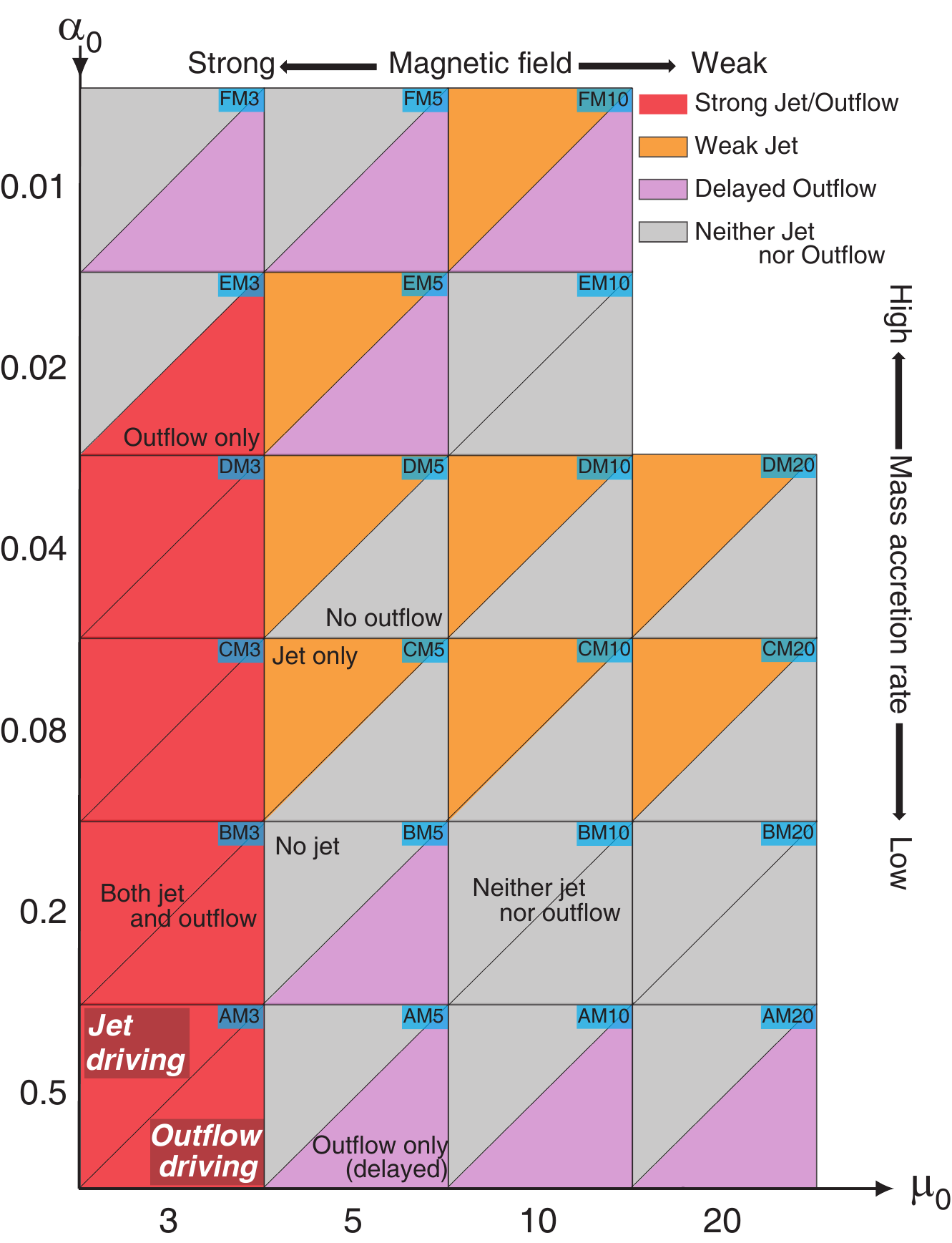}
\end{center}
\caption{
Outcomes of this study and Paper III plotted on the parameter plane ($\mu_0$ and $\alpha_0$).
The model name is described in each panel. 
In each panel, the outcomes of this study (i.e. jet driving) and that of Paper III are represented by color in the upper left and the lower right, respectively. 
Color is used to indicate strong jets (upper left) or strong outflow (lower right) (red),  no jet (this study, upper left) or no outflow (Paper III, lower right) (gray),  a weak jet having a low momentum in this study (orange) and  an outflow in a delayed fashion in Paper III (purple).  
}
\label{fig:14}
\end{figure*}

Figure~\ref{fig:14} shows the outcomes of this study and Paper III. 
In each panel, the outcome of this study is presented in the upper left, while that of Paper III is in the lower right.
The model names used in this study are the same as those in Paper III when the parameters $\alpha_0$  and $\mu_0$ are the same.  
The initial condition is almost the same between the two studies.
The difference in the initial condition is the angle between the magnetic and angular momentum vectors of the prestellar cloud $\theta_0$, as described in \S\ref{sec:settings}.  
The angle $\theta_0=0$ was adopted in Paper III, while $\theta_0=20^\circ$ is used in this study.  
Thus, the initial magnetic field  and angular momentum vector are aligned in Paper III, while they are misaligned in this study. 
In addition,  the integration time $t_{\rm end}$ and spatial resolution $h$ are different between this study ($t_{\rm end}\sim500$\,yr and $h=7.9\times10^{-3}$\,au ) and Paper III ($t_{\rm end}\sim10^4$\,yr and $h=0.62$\,au).
%% there exist differences in the treatment of sink cells and the integration time in the simulations. 
%%The calculations were executed for $\sim10^4$\,yr after protostar formation with sink cells in Paper III, while for $\sim500$ without sink cells in this study (see \S\ref{sec:settings}).
%%Thus, I cannot fairly compare the models between Paper III and this study. 
Hence, it is difficult to fairly compare the outcomes between them. 
However,  we cannot realize a long-term cloud evolution with a high-spatial resolution.
Thus,  it would be valuable to overview the results between them in order to further understand the driving mechanism of low- and high-velocity flows.
%%Thus, we  use the results in Paper III and this study only for discussion. 
In this section, I discuss the possibility of jet and outflow driving in the clouds with different magnetic field strengths and mass accretion rates using the results in Paper III and this study.

In \S\ref{sec:results}, all the outflowing gas observed in this study is referred to as outflow, with  the outflow being composed of both high-velocity (or high-velocity jet) and low-velocity (low-velocity outflow) components.
Here, I redefine the terms  `jet' and `outflow', {\it used only in the following paragraphs in this section}, to distinguish the terms used for the outflow between this study and  Paper III.  
In this section, I refer to the outflow as the outgoing flow observed in {\it Paper III}, while I call the jet the outgoing flows observed {\it in this study}. 
In addition, referring to Figures~\ref{fig:11}--\ref{fig:13}, I call a jet with a large momentum a strong jet and one with small momentum a weak jet.  
Furthermore, in this study, since the low-velocity component appears with the high-velocity jet (or high-velocity component),  as seen in Figure~\ref{fig:8}, I call the low-velocity component of  a jet  the low-velocity outflow (or low-velocity component) and the high-velocity flow the high-velocity jet  (high-velocity component).
As seen in Figure~\ref{fig:8},  there is no clear boundary between a high-velocity jet and low-velocity outflow.
Thus, it is difficult to separate low-velocity outflow from a high-velocity jet. 
The low-velocity outflow may be directly driven by the disk outer edge, as explain in Papers I--III.
Alternatively, it may be entrained by the high-velocity component when the high-velocity jet appears. 
In the present study, I do not consider the driving mechanism of the low-velocity components appearing in this study, as this is outside the scope of this study.

First, I focus on the models with $\alpha_0=0.2$ and 0.5 (the fifth and sixth lines in Fig.~\ref{fig:14}; models AM3, AM5, AM10, AM20, BM3, BM5, BM10, BM20), which have a low mass accretion rate (Fig.~\ref{fig:10}) and would form low mass stars. 
Figure~\ref{fig:14} shows that both outflow and jets appear only in AM3 and BM3, which have the  strongest magnetic fields ($\mu_0=3$). 
For these models, the outflow appears  without resolving the jet driving region (Paper III). 
The  jet also appears as described in \S\ref{sec:results}.
Thus, it is considered that a high-velocity jet and low-velocity outflow are driven at different radii in the circumstellar disk, as expected from past core collapse simulations \citep{tomida13,machida19}. 
On the other hand, when the magnetic field is not strong ($\mu_0 \ge 5$; AM5, AM10, AM20, BM5, BM10, BM20), jets do not appear.
In Figure~\ref{fig:14}, `Delayed Outflow' refers to the case where a (weak) outflow begins to grow only in the late accretion phase, just before the infalling envelope dissipates (Paper III).
Also, for BM10 and BM20, neither jets (this study) nor outflow (Paper III) appear. 
Thus, for these models, the outflow and jet  would not  play a significant role by  the end of the mass accretion phase.  
Note that since the jet driving was calculated only for $\sim500$\,yr in the main accretion phase with a high spatial resolution,  I cannot clearly state that the jet driving is completely suppressed  by the end of the mass accretion phase. 
However, since no jet appears, it cannot help to drive the low-velocity outflow, at least, in the early main accretion phase. 

Next, I describe the models with $\alpha_0=0.04$ and 0.08 (the third and fourth lines in Fig.~\ref{fig:14}; CM3, CM5, CM10, CM20, DM3, DM5, DM10, DM20), which have moderate mass accretion rates (Fig.~\ref{fig:11}) and would form intermediate mass stars.
Even for these models, both the jet and outflow appear only when the magnetic field is as strong as $\mu_0=3$. 
Thus, as described above, the jet and outflow are expected to be driven at different radii of the circumstellar disk. 
The difference between low-mass ($\alpha_0=0.2$ and 0.5) and intermediate-mass ($\alpha=0.04$ and 0.08) stars can be seen in the models with weak magnetic fields. 
When the magnetic field is as weak as $\mu\ge 5$, neither jets nor outflow appear in the early accretion phase for the low-mass star formation cases.
On the other hand, a weak jet appears for all models with $\mu_0 \ge 5$ (CM5, CM10, CM20, DM5, DM10, DM20) for intermediate star formation cases.  
As shown in Figure~\ref{fig:8}, in addition to the high-velocity component, the low-velocity component exists when a jet  appears  (Fig.~\ref{fig:8}).
Thus, a part of the outflowing gas would be entrained by the high-velocity jet. 
Alternatively, the jet may alleviate the ram pressure barrier which suppresses the growth  of the low-velocity outflow and promotes outflow directly driven by the outer disk region. 
However, the momentum of the outflowing gas in these weak jet models is not as large as that in strong jet models (Fig.~\ref{fig:12}).  
Thus, the contribution of the jet to the low-velocity outflow is  limited. 

Finally, I describe the models with $\alpha_0 = 0.01$ and 0.02 (the first and second lines in Fig.~\ref{fig:14}; EM3, EM5, EM10, FM3, DM5, FM10) which have high mass accretion rates (Fig.~\ref{fig:11}) and would form high mass stars. 
No strong jets appear in these models, while only EM3 shows an outflow after protostar formation. 
Model EM3 is a special case of all the models listed in Table~\ref{table:1}, because only this model shows a low-velocity outflow (Paper III) without showing a jet (this study). 
In Paper III, the  outflow begins to grow about 1,000\,yr after protostar formation in model EM3. 
Thus,  the protostellar system drives only the low-velocity outflow without the high-velocity component, which may explain the observation of \citet{hirota17}. 

For EM5 and FM10, a weak jet appears in the early mass accretion phase (this study), while the outflow appears in the later accretion phase (Paper III). 
Meanwhile, for FM3 and FM5, the outflow appears in the late accretion phase  (Paper III), while no jet appears (this study). 
%%before protostar formation 
For EM10, neither jet nor outflow appear in either study. 
Thus, it is expected that, for these models (EM5, EM10, FM3, FM5, FM10), jets and outflow do not significantly affect the star formation process throughout the main accretion phase. 
In addition, in the models with $\alpha_0 = 0.01$ and 0.02, jets would not greatly contribute to driving the low-velocity outflow because of a lack of strong jets.

The protostellar outflow is considered to have a significant impact on star formation. 
However, Figure~\ref{fig:14} indicates that there are various star formation scenarios (with and without jets and outflow) dependent on the initial prestellar cloud's magnetic field strength and intrinsic mass accretion rate.

Finally,  I again emphasize the caveat in the above discussion. 
In Papers I-III, I prepared the aligned models, in which the rotation axis of the prestellar cloud is aligned with the global magnetic field. 
Since such a setting is not very realistic, I prepared the misaligned models in this study (see \S\ref{sec:settings}). 
In addition,  although  the protostar was spatially resolved, I could not follow the long-term evolution of the jet.
For further understanding of the  jet and outflow driving, future studies should  focus on the long-term evolution of the jet resolving the protostar with the parameters $\mu_0$, $\alpha_0$ and $\theta_0$.

\section{Summary} 
In this study, I investigated jet driving in clouds with different magnetic field strengths and intrinsic mass accretion rates using numerical simulations. 
I found that a high-velocity jet appears near the protostar and is sustained for a long time only when the magnetic field strength of the prestellar cloud is as strong as $\mu_0\le3$ and the mass accretion rate onto the central region is as low as  $\dot{M} < 10^{-4}\,\msun$\,yr$^{-1}$. 
When strongly magnetized clouds ($\mu_0=3$) have  a mass accretion rate higher than  $\dot{M} > 10^{-4}\,\msun$\,yr$^{-1}$,  a high-velocity jet can be driven near the protostar in the early mass accretion phase.
However, the jet does not grow sufficiently and quickly disappears. 
On the other hand, when the magnetic field of the star-forming cloud is weak,  a jet sometimes appears for a mass accretion rate higher than  $\dot{M} > 10^{-5}\,\msun$\,yr$^{-1}$.
In this case, a low-velocity component also appears in conjunction with the high-velocity jet. 
However, the outflow momentum in weakly magnetized clouds is much lower than that in strongly magnetized clouds.  
Thus, a high-velocity jet does not significantly contribute to driving a low-velocity outflow.
This study indicates that  a protostellar outflow cannot significantly contribute to the star formation process when the magnetic field strength of the star-forming cloud is weak. 
%% results of this and previous studies (Paper III)

\section*{Acknowledgements}
This research used the computational resources of the HPCI system provided by the Cyber Science Center at Tohoku University and the Cybermedia Center at Osaka University (Project ID: hp200004, hp210004).
Simulations reported in this paper were also performed by 2019 and 2020 Koubo Kadai on Earth Simulator (NEC SX-ACE) at JAMSTEC. 
The present study was supported by JSPS KAKENHI Grant (JP17H06360, JP17K0538faa7, JP17KK0096, JP21H00046, JP21K03617: MNM).

\section*{DATA AVAILABILITY}
The data underlying this article are available in the article and in its
online supplementary material.

%%\bsp	% typesetting comment
%%\label{lastpage}

\begin{thebibliography}{}{}
\bibitem[\protect\citeauthoryear{Alves et al.}{2017}]{alves17} 
Alves F.~O., Girart J.~M., Caselli P., Franco G.~A.~P., Zhao B., Vlemmings W.~H.~T., Evans M.~G., et al., 2017, A\&A, 603, L3. doi:10.1051/0004-6361/201731077

\bibitem[\protect\citeauthoryear{Aso et al.}{2019}]{aso19} 
Aso Y., Hirano N., Aikawa Y., Machida M.~N., Ohashi N., Saito M., Takakuwa S., et al., 2019, ApJ, 887, 209. doi:10.3847/1538-4357/ab5284

\bibitem[\protect\citeauthoryear{Arce et al.}{2007}]{arce07} 
Arce H.~G., Shepherd D., Gueth F., Lee C.-F., Bachiller R., Rosen A., Beuther H., 2007, prpl.conf, 245

\bibitem[\protect\citeauthoryear{Bate, Tricco, \& Price}{2014}]{bate14} 
Bate M.~R., Tricco T.~S., Price D.~J., 2014, MNRAS, 437, 77. doi:10.1093/mnras/stt1865

\bibitem[\protect\citeauthoryear{Bjerkeli et al.}{2016}]{bjerkeli16} 
Bjerkeli P., van der Wiel M.~H.~D., Harsono D., Ramsey J.~P., J{\o}rgensen J.~K., 2016, Natur, 540, 406. doi:10.1038/nature20600

%%\bibitem[Bhandare et al.(2018)]{bhandare18} 
%%Bhandare, A., Kuiper, R., Henning, T., et al.\ 2018, A\&A, 618, A95

\bibitem[\protect\citeauthoryear{Banerjee \& Pudritz}{2006}]{banerjee06} 
Banerjee R., Pudritz R.~E., 2006, ApJ, 641, 949. doi:10.1086/500496

\bibitem[\protect\citeauthoryear{Blandford \& Payne}{1982}]{blandford82} 
Blandford R.~D., Payne D.~G., 1982, MNRAS, 199, 883. doi:10.1093/mnras/199.4.883

\bibitem[\protect\citeauthoryear{Ciardi \& Hennebelle}{2010}]{ciardi10} 
Ciardi A., Hennebelle P., 2010, MNRAS, 409, L39. doi:10.1111/j.1745-3933.2010.00942.x

%%\bibitem[\protect\citeauthoryear{Commer{\c{c}}on, Hennebelle, \& Henning}{2011}]{commercon11} 
%%Commer{\c{c}}on B., Hennebelle P., Henning T., 2011, ApJL, 742, L9. doi:10.1088/2041-8205/742/1/L9

\bibitem[\protect\citeauthoryear{de Valon et al.}{2020}]{devalon20} 
de Valon A., Dougados C., Cabrit S., Louvet F., Zapata L.~A., Mardones D., 2020, A\&A, 634, L12. doi:10.1051/0004-6361/201936950


\bibitem[\protect\citeauthoryear{Hennebelle \& Fromang}{2008}]{hennebelle08} 
Hennebelle P., Fromang S., 2008, A\&A, 477, 9. doi:10.1051/0004-6361:20078309

\bibitem[\protect\citeauthoryear{Hirano \& Machida}{2019}]{hirano19} 
Hirano S., Machida M.~N., 2019, MNRAS, 485, 4667. doi:10.1093/mnras/stz740

\bibitem[\protect\citeauthoryear{Hirano et al.}{2020}]{hirano20} 
Hirano S., Tsukamoto Y., Basu S., Machida M.~N., 2020, ApJ, 898, 118. doi:10.3847/1538-4357/ab9f9d

\bibitem[\protect\citeauthoryear{Hirota et al.}{2017}]{hirota17} 
Hirota T., Machida M.~N., Matsushita Y., Motogi K., Matsumoto N., Kim M.~K., Burns R.~A., et al., 2017, NatAs, 1, 0146. doi:10.1038/s41550-017-0146

\bibitem[\protect\citeauthoryear{Inutsuka}{2012}]{inutsuka12} 
Inutsuka S., 2012, PTEP, 2012, 01A307. doi:10.1093/ptep/pts024

\bibitem[\protect\citeauthoryear{Joos, Hennebelle, \& Ciardi}{2012}]{joos12} 
Joos M., Hennebelle P., Ciardi A., 2012, A\&A, 543, A128. doi:10.1051/0004-6361/201118730

\bibitem[\protect\citeauthoryear{K{\"o}lligan \& Kuiper}{2018}]{kolligan18} 
K{\"o}lligan A., Kuiper R., 2018, A\&A, 620, A182. doi:10.1051/0004-6361/201833686

\bibitem[\protect\citeauthoryear{Lee et al.}{2021}]{lee21} 
Lee C.-F., Tabone B., Cabrit S., Codella C., Podio L., Ferreira J., Jacquemin-Ide J., 2021, ApJL, 907, L41. doi:10.3847/2041-8213/abda38

\bibitem[\protect\citeauthoryear{Lewis, Bate, \& Price}{2015}]{lewis15} Lewis B.~T., Bate M.~R., Price D.~J., 2015, MNRAS, 451, 288. doi:10.1093/mnras/stv957

\bibitem[\protect\citeauthoryear{Lewis \& Bate}{2017}]{lewis17} 
Lewis B.~T., Bate M.~R., 2017, MNRAS, 467, 3324. doi:10.1093/mnras/stx271

\bibitem[\protect\citeauthoryear{Marchand et al.}{2020}]{marchand20} 
Marchand P., Tomida K., Tanaka K.~E.~I., Commer{\c{c}}on B., Chabrier G., 2020, ApJ, 900, 180. doi:10.3847/1538-4357/abad99


\bibitem[\protect\citeauthoryear{Machida et al.}{2004}]{machida04} 
 Machida, M. N., Tomisaka, K., \& Matsumoto, T.\ 2004, MNRAS, 348, L1 

\bibitem[\protect\citeauthoryear{Machida \etal}{2005a}]{machida05a}
 Machida, M. N., Matsumoto, T., Tomisaka, K., \& Hanawa, T. 2005, MNRAS, 362, 369

\bibitem[\protect\citeauthoryear{Machida, Inutsuka, \& Matsumoto}{2006}]{machida06} 
Machida M.~N., Inutsuka S., Matsumoto T., 2006, ApJL, 647, L151. doi:10.1086/507179

\bibitem[\protect\citeauthoryear{Machida, Inutsuka, \& Matsumoto}{2007}]{machida07} 
Machida M.~N., Inutsuka S., Matsumoto T., 2007, ApJ, 670, 1198. doi:10.1086/521779

\bibitem[\protect\citeauthoryear{Machida, Inutsuka, \& Matsumoto}{2008}]{machida08} 
Machida M.~N., Inutsuka S., Matsumoto T., 2008, ApJ, 676, 1088. doi:10.1086/528364

\bibitem[Machida et al.(2010)]{machida10} 
Machida, M.~N., Inutsuka, S., \& Matsumoto, T.\ 2010, \apj, 724, 1006

%%\bibitem[Machida et al.(2011)]{machida11b} 
%%Machida, M.~N., Inutsuka, S., \& Matsumoto, T.\ 2011, \pasj, 63, 555

%%\bibitem[Machida \& Matsumoto(2011)]{machida11} 
%%Machida, M.~N., \& Matsumoto, T.\ 2011, \mnras, 413, 2767

\bibitem[\protect\citeauthoryear{Machida \& Hosokawa}{2013}]{machida13} 
Machida M.~N., Hosokawa T., 2013, MNRAS, 431, 1719. doi:10.1093/mnras/stt291

\bibitem[\protect\citeauthoryear{Machida}{2014}]{machida14} 
Machida M.~N., 2014, ApJL, 796, L17. doi:10.1088/2041-8205/796/1/L17

\bibitem[\protect\citeauthoryear{Machida \& Nakamura}{2015}]{machida15} 
Machida M.~N., Nakamura T., 2015, MNRAS, 448, 1405. doi:10.1093/mnras/stu2633

\bibitem[\protect\citeauthoryear{Machida \& Basu}{2019}]{machida19} 
Machida M.~N., Basu S., 2019, ApJ, 876, 149. doi:10.3847/1538-4357/ab18a7

\bibitem[\protect\citeauthoryear{Machida \& Hosokawa}{2020}]{machida20} 
Machida M.~N., Hosokawa T., 2020a, MNRAS, 499, 4490. doi:10.1093/mnras/staa3139

\bibitem[\protect\citeauthoryear{Machida, Hirano, \& Kitta}{2020}]{machida20b} 
Machida M.~N., Hirano S., Kitta H., 2020b, MNRAS, 491, 2180. doi:10.1093/mnras/stz3159

\bibitem[\protect\citeauthoryear{Matsushita et al.}{2017}]{matsushita17} 
Matsushita Y., Machida M.~N., Sakurai Y., Hosokawa T., 2017, MNRAS, 470, 1026. doi:10.1093/mnras/stx893

\bibitem[\protect\citeauthoryear{Matsushita et al.}{2018}]{matsushita18} 
Matsushita Y., Sakurai Y., Hosokawa T., Machida M.~N., 2018, MNRAS, 475, 391. doi:10.1093/mnras/stx3070

\bibitem[\protect\citeauthoryear{Matsushita et al.}{2019}]{matsushita19} 
Matsushita Y., Takahashi S., Machida M.~N., Tomisaka K., 2019, ApJ, 871, 221. doi:10.3847/1538-4357/aaf1b6

%%\bibitem[\protect\citeauthoryear{Matzner \& McKee}{2000}]{matzner00} 
%%Matzner C.~D., McKee C.~F., 2000, ApJ, 545, 364. doi:10.1086/317785

\bibitem[\protect\citeauthoryear{Masson et al.}{2016}]{masson16} 
Masson J., Chabrier G., Hennebelle P., Vaytet N., Commer{\c{c}}on B., 2016, A\&A, 587, A32. doi:10.1051/0004-6361/201526371

\bibitem[\protect\citeauthoryear{Matzner \& McKee}{2000}]{matzner00} 
Matzner C.~D., McKee C.~F., 2000, ApJ, 545, 364. doi:10.1086/317785

\bibitem[\protect\citeauthoryear{Saiki \& Machida}{2020}]{saiki20} 
Saiki Y., Machida M.~N., 2020, ApJL, 897, L22. doi:10.3847/2041-8213/ab9d86

\bibitem[\protect\citeauthoryear{Seifried et al.}{2011}]{seifried11} 
Seifried D., Banerjee R., Klessen R.~S., Duffin D., Pudritz R.~E., 2011, MNRAS, 417, 1054. doi:10.1111/j.1365-2966.2011.19320.x

\bibitem[Seifried et al.(2012)]{seifried12} 
Seifried, D., Pudritz, R.~E., Banerjee, R., Duffin, D., \& Klessen, R.~S.\ 2012, MNRAS, 422, 347 

\bibitem[\protect\citeauthoryear{Tabone et al.}{2018}]{tabone18} 
Tabone B., Raga A., Cabrit S., Pineau des For{\^e}ts G., 2018, A\&A, 614, A119. doi:10.1051/0004-6361/201732031

\bibitem[Tanaka et al.(2017)]{tanaka17} 
Tanaka, K.~E.~I., Tan, J.~C., \& Zhang, Y.\ 2017, \apj, 835, 32

\bibitem[Tanaka et al.(2018)]{tanaka18} 
Tanaka, K.~E.~I., Tan, J.~C., Zhang, Y., et al.\ 2018, \apj, 861, 68

%%\bibitem[Tan et al.(2013)]{tan13} 
%%Tan, J.~C., Kong, S., Butler, M.~J., Caselli, P., \& Fontani, F.\ 2013, ApJ, 779, 96 

%%\bibitem[Tan et al.(2014)]{tan14} 
%%Tan, J.~C., Beltr{\'a}n, M.~T., Caselli, P., et al.\ 2014, Protostars and Planets VI, 149

%%\bibitem[Tan et al.(2016)]{tan16} 
%%Tan, J.~C., Kong, S., Zhang, Y., et al.\ 2016, \apjl, 821, L3

\bibitem[\protect\citeauthoryear{Tokuda et al.}{2018}]{tokuda18} 
Tokuda K., Onishi T., Saigo K., Matsumoto T., Inoue T., Inutsuka S., Fukui Y., et al., 2018, ApJ, 862, 8. doi:10.3847/1538-4357/aac898

\bibitem[\protect\citeauthoryear{Tomida et al.}{2013}]{tomida13} 
Tomida K., Tomisaka K., Matsumoto T., Hori Y., Okuzumi S., Machida M.~N., Saigo K., 2013, ApJ, 763, 6. doi:10.1088/0004-637X/763/1/6

%%\bibitem[Tomisaka(1998)]{tomisaka98} 
%%Tomisaka, K.\ 1998, ApJL, 502, L163 

\bibitem[\protect\citeauthoryear{Tomisaka}{2000}]{tomisaka00} 
Tomisaka K., 2000, ApJL, 528, L41. doi:10.1086/312417

\bibitem[\protect\citeauthoryear{Tomisaka}{2002}]{tomisaka02} 
Tomisaka K., 2002, ApJ, 575, 306. doi:10.1086/341133

%%\bibitem[Toomre(1964)]{toomre64} 
%% Toomre, A.\ 1964, ApJ, 139, 1217 

%%\bibitem[\protect\citeauthoryear{Troland \& Crutcher}{2008}]{troland08} 
%%Troland T.~H., Crutcher R.~M., 2008, ApJ, 680, 457

\bibitem[Truelove et al.(1997)]{truelove97} 
Truelove, J.~K., Klein, R.~I., McKee, C.~F., et al.\ 1997, \apjl, 489, L179

\bibitem[\protect\citeauthoryear{Uchida \& Shibata}{1985}]{uchida85} 
Uchida Y., Shibata K., 1985, PASJ, 37, 515

\bibitem[\protect\citeauthoryear{Vaytet et al.}{2018}]{vaytet18} 
Vaytet N., Commer{\c{c}}on B., Masson J., Gonz{\'a}lez M., Chabrier G., 2018, A\&A, 615, A5. doi:10.1051/0004-6361/201732075

\bibitem[\protect\citeauthoryear{Wurster, Price, \& Bate}{2016}]{wurster16} 
Wurster J., Price D.~J., Bate M.~R., 2016, MNRAS, 457, 1037. doi:10.1093/mnras/stw013

\bibitem[\protect\citeauthoryear{Wurster, Bate, \& Price}{2018}]{wurster18} 
Wurster J., Bate M.~R., Price D.~J., 2018, MNRAS, 475, 1859. doi:10.1093/mnras/stx3339

\bibitem[\protect\citeauthoryear{Wurster, Bate, \& Bonnell}{2021}]{wurster21} 
Wurster J., Bate M.~R., Bonnell I.~A., 2021, MNRAS.tmp. doi:10.1093/mnras/stab2296


\bibitem[\protect\citeauthoryear{Xu \& Kunz}{2021}]{xu21} 
Xu W., Kunz M.~W., 2021, MNRAS, 502, 4911. doi:10.1093/mnras/stab314


\bibitem[\protect\citeauthoryear{Wu et al.}{2004}]{wu04} 
Wu Y., Wei Y., Zhao M., Shi Y., Yu W., Qin S., Huang M., 2004, A\&A, 426, 503. doi:10.1051/0004-6361:20035767

%%\bibitem[Zhang et al.(2005)]{zhang05} 
%%Zhang, Q., Hunter, T.~R., Brand, J., et al.\ 2005, ApJ, 625, 864 

%%\bibitem[Zhang et al.(2009)]{zhang09} 
%%Zhang, Q., Wang, Y., Pillai, T., et al.\ 2009, \apj, 696, 268

%%\bibitem[Zhang et al.(2014)]{zhang14} 
%%Zhang, Q., Qiu, K., Girart, J.~M., et al.\ 2014, ApJ, 792, 116 
\end{thebibliography}
\end{document}